\PassOptionsToPackage{unicode}{hyperref}
\PassOptionsToPackage{hyphens}{url}
\documentclass[
]{article}
\usepackage{lmodern}
\usepackage{amssymb,amsmath}
\usepackage{ifxetex,ifluatex}
\ifnum 0\ifxetex 1\fi\ifluatex 1\fi=0 
\usepackage[T1]{fontenc}
\usepackage[utf8]{inputenc}
\usepackage{textcomp} 
\else 
\usepackage{unicode-math}
\defaultfontfeatures{Scale=MatchLowercase}
\defaultfontfeatures[\rmfamily]{Ligatures=TeX,Scale=1}
\fi
\IfFileExists{upquote.sty}{\usepackage{upquote}}{}
\IfFileExists{microtype.sty}{
	\usepackage[]{microtype}
	\UseMicrotypeSet[protrusion]{basicmath} 
}{}
\makeatletter
\@ifundefined{KOMAClassName}{
	\IfFileExists{parskip.sty}{%
		\usepackage{parskip}
	}{
		\setlength{\parindent}{0pt}
		\setlength{\parskip}{6pt plus 2pt minus 1pt}}
}{
	\KOMAoptions{parskip=half}}
\makeatother
\usepackage{xcolor}
\IfFileExists{xurl.sty}{\usepackage{xurl}}{} 
\IfFileExists{bookmark.sty}{\usepackage{bookmark}}{\usepackage{hyperref}}
\hypersetup{
	pdftitle={MPA network design based on graph theory and emergent properties of larval dispersal},
	hidelinks,
	pdfcreator={LaTeX via pandoc}}
\usepackage{longtable,booktabs}
\usepackage{etoolbox}
\makeatletter
\patchcmd\longtable{\par}{\if@noskipsec\mbox{}\fi\par}{}{}
\makeatother
\IfFileExists{footnotehyper.sty}{\usepackage{footnotehyper}}{\usepackage{footnote}}
\makesavenoteenv{longtable}
\usepackage{graphicx}
\usepackage{hyperref}
\usepackage{pdflscape} 
\usepackage{multirow}
\usepackage{setspace}
\usepackage{lineno}
\usepackage{authblk}
\usepackage{soul}

\usepackage{pdfpages}

\usepackage{fancyhdr}

\title{MPA network design based on graph theory and emergent properties of
	larval dispersal\footnote{© 2020. This manuscript version is made available under the CC-BY-NC-ND 4.0 license \href{http://creativecommons.org/licenses/by-nc-nd/4.0/}{http://creativecommons.org/licenses/by-nc-nd/4.0/}}}

\author[1]{Andres Ospina-Alvarez\footnote{Corresponding author: Andrés Ospina-Alvarez, email: aospina.co@me.com; address: Spanish Scientific Research Council, Mediterranean Institute for Advanced Studies (IMEDEA-CSIC/UIB), C/ Miquel Marques 21, CP 07190 Esporles, Balearic Islands, Spain.}}

\author[2]{Silvia de Juan Mohan}

\author[2]{Josep Alós }

\author[2]{Gotzon Basterretxea}

\author[3]{Alexandre Alonso-Fernández}

\author[2]{Guillermo Follana-Berná}

\author[2]{Miquel Palmer}

\author[2]{Ignacio A. Catalán}

\affil[1]{Mediterranean Institute for Advanced Studies (IMEDEA-CSIC/UIB), C/
	Miquel Marques 21, CP 07190 Esporles, Balearic Islands, Spain.}
\affil[2]{Marine Science Institute (ICM-CSIC), Passeig Marítim de la
	Barceloneta, 37-49, CP 08003 Barcelona, Catalunya, Spain.}
\affil[3]{Marine Research Institute (IIM-CSIC). C/ Eduardo Cabello 6, CP 36208
	Vigo, Pontevedra, Spain.}

\date{}

\begin{document}
	\maketitle
	
	\textbf{Running page head:} Graph theory for designing MPA networks
	
	\begin{abstract}

Despite the recognised effectiveness of networks of Marine Protected
Areas (MPAs) as a biodiversity conservation instrument, nowadays MPA
network design frequently disregards the importance of connectivity
patterns. In the case of sedentary marine populations, connectivity
stems not only from the stochastic nature of the physical environment
that affects early-life stages dispersal, but also from the spawning
stock attributes that affect the reproductive output (e.g., passive eggs
and larvae) and its survivorship. Early-life stages are virtually
impossible to track in the ocean. Therefore, numerical ocean current
simulations coupled to egg and larval Lagrangian transport models remain
the most common approach for the assessment of marine larval
connectivity. Inferred larval connectivity may be different depending on
the type of connectivity considered; consequently, the prioritisation of
sites for marine populations' conservation might also differ. Here, we
introduce a framework for evaluating and designing MPA networks based on
the identification of connectivity hotspots using graph theoretic
analysis. We use as a case of study a network of open-access areas and
MPAs, off Mallorca Island (Spain), and test its effectiveness for the
protection of the painted comber \emph{Serranus scriba}. Outputs from
network analysis are used to: (1) identify critical areas for improving
overall larval connectivity; (2) assess the impact of species'
biological parameters in network connectivity; and (3) explore
alternative MPA configurations to improve average network connectivity.
Results demonstrate the potential of graph theory to identify
non-trivial egg/larval dispersal patterns and emerging collective
properties of the MPA network which are relevant for increasing
protection efficiency.

\end{abstract}


\section{Introduction}

In ecology, connectivity concerns the exchange of individuals among
populations and, thereby, the dispersal ability and the spatial and
temporal scales over which a population of a given species is connected
with other populations {(Strathmann et al. 2002, Armsworth 2002)}. Most
marine species have a pelagic early-life stage in which individuals
(i.e., eggs, spores, larvae, juveniles) drift away from natal locations
transported by ocean currents. For many species, these early life stages
represent the most important, or even the only mechanism of dispersal
{(Walford 1938, Norcross \& Shaw 1984)}. Although marine connectivity
often refers only to a part of the ``reproductive resilience''
{(Lowerre-Barbieri et al. 2016)}, it remains a key component of the
system affecting the recruitment and, as a consequence, the long-term
population dynamics and persistence {(Travis \& Dytham 1998, Siegel et
al. 2003, Watson et al. 2012)}. As such, connectivity is of great
interest for the management of marine ecosystems, and a key variable to
consider in the scale, spacing and spatial structure of Marine Protected
Areas (MPAs) {(Lubchenco et al. 2003, Botsford et al. 2003, Fogarty \&
Botsford 2007, Botsford et al. 2009)}.

The identification of population connectivity patterns is, however,
challenging since dispersal of early-life stages is highly influenced by
the stochastic nature of the physical environment {(Siegel et al. 2008)
}and by multiple biological factors affecting the reproductive output of
the spawning stock biomass (SSB), like timing of spawning, egg
physiology, Pelagic Larval Duration (PLD), larval behaviour, and/or
larval mortality {(Hinckley et al. 2001, Galarza et al. 2009,
Ospina-Alvarez, Palomera, et al. 2012, Morgan 2014, Donahue et al. 2015,
Ospina-Alvarez et al. 2018)}. While displacement routes of large-size
organisms (i.e., adult fish) can be assessed using individual-tracking
devices, planktonic individuals such as eggs and larvae are virtually
impossible to track in the ocean {(but see, Paris, Helgers, et al.
2013)}, notably limiting our ability to characterise their dispersal
{(Pineda 2000, Armsworth 2002, Siegel et al. 2003, Kinlan \& Gaines
2003, Cowen et al. 2006, Cowen \& Sponaugle 2009)}. This limitation has
encouraged the use of a suite of indirect methods to reconstruct
probable dispersal pathways, such as the analysis of geochemical
signatures in calcified structures {(Thorrold et al. 2007)} and genetic
structure assessments {(Hedgecock et al. 2007)}. However, numerical
ocean current simulations coupled to egg and larval Lagrangian transport
algorithms (i.e., Individual Based Models IBMs) remain the most common
approach for the assessment of marine larval connectivity {(Werner et
al. 2007, Catalán et al. 2013, Alós et al. 2014, Ospina-Alvarez et al.
2015, Blanco et al. 2019)}.

Larval connectivity derived from these IBMs can be interpreted using two
measures: potential and realised larval connectivity{ (Watson et al.
2010)}. Potential larval connectivity refers to the probability of
connection between a natal site or spawning area and a destination or
nursery area. In contrast, realised larval connectivity refers to the
number of larvae traveling from spawning to nursery sites {(Watson et
al. 2010)}. Realised larval connectivity can be estimated using
potential connectivity weighted by relevant biological and environmental
information {(Kough \& Paris 2015)}. For example, an estimate of
realised connectivity can be obtained by weighting potential larval
connectivity with observed or modelled spatial egg production {(see
review by Lowerre-Barbieri et al. 2016)}. These estimates can be
improved by including biological aspects such as maternal effects (like
body size), reproductive timing, spawning seasonality or individual
spawning times {(Hixon et al. 2014, Gwinn et al. 2015)}. Realised larval
connectivity provides spatially explicit information on the ability of a
set of sites to hold connected populations. Therefore, it can provide
valuable information for the management of coastal resources
{(Ospina-Alvarez et al. 2020)}. However, it is rarely considered in the
design of MPA networks {(but see Watson et al. 2010)}.

The effectiveness of an MPA network critically depends upon the
consideration of connectivity patterns, source-sink dynamics and rates
of population replenishment {(Botsford et al. 2008, 2009, Gaines et al.
2010, Lagabrielle et al. 2014)}. Therefore, connectivity values
estimated trough IBM can provide insights into the design of MPA
networks. However, conventional connectivity matrices generated by
larval transport IBMs only provide a partial representation of the
complexity of the connectivity and omit parts such as the identification
of key locations that act as genetic corridors or central populations
that feed many others. In this context, graph theory (i.e., the
mathematical study of the interaction of a system of connected elements)
is a valuable approach for analysing MPA network performance, as it
provides a simplified and quantitative view of the multiple factors
involved in the exchange among system elements {(Conklin et al. 2018,
Henry et al. 2018, Kininmonth et al. 2018, Friesen et al. 2019)}.{~}

In graph theory, a system of connected elements can be defined as a
"network", also called a "graph''. Network elements are modelled as
``vertices'' or ``nodes'' in the graph and their connections or links
are represented as ``edges'' or ``arcs''. In a network of marine
reserves, the graph represents the network itself, with each area being
a vertex or node, and the probability of connection or flow of
individuals between areas being the arcs or edges {(Dale \& Fortin
2010)}. Graph theory provides insights into the system properties and
identifies critical nodes with high centrality (i.e., connected to many
other areas) or clusters of well-connected nodes with high potential
genetic flow and acting as bridges between distant populations {(Treml
et al. 2008, Kininmonth et al. 2009, Jacobi \& Jonsson 2011, Friesen et
al. 2019)}. MPA network design based on connectivity studies and graph
theory have increased in recent years {(Treml et al. 2008, Kininmonth et
al. 2018, Friesen et al. 2019)}. However, most studies fail to adopt an
objective approach for selecting the most adequate centrality measures
to identify important nodes in an MPA network. The difficulty lies in
the fact that each node could be important from a different point of
view depending on the definition of "importance". According to Freeman
{(1978)}: "There is certainly no unanimity on exactly what centrality
is, or on its conceptual foundations, and there is little agreement on
the appropriate procedure for its measurement''.

The methodological framework introduced in this study allows addressing
several questions: 1) Does the importance of sites for population
connectivity differ when demographic characteristics and available
habitat are coupled to larval transport models? 2) Does the importance
of sites for population connectivity differ when considering either a)
the probability or flow of individuals and b) centrality measures from
graph theoretic analyses? 3) Is it possible to design an MPA network
that maximises net larval productivity, net larval supply and,
simultaneously, avoids the fragmentation between sub-populations? For
this purpose, we first analyse the potential and realised connectivity
and the emergent properties of an ensemble of coastal species inhabiting
an MPA network, where the respective sub-populations are potentially
interconnected (e.g., in terms of dispersive/retentive associations
between areas) through early-life stages {(Basterretxea et al. 2012)}.
As a novelty, our approach explores differences between the potential
and realised networks and provides an ecological interpretation to the
centrality measures analysed. This approach sets the baseline for a
wider adoption of coupled connectivity-graph theory models in marine
conservation and planning.

\section{Methods}

\begin{figure}
	\centering
	\includegraphics[width=0.9\linewidth]{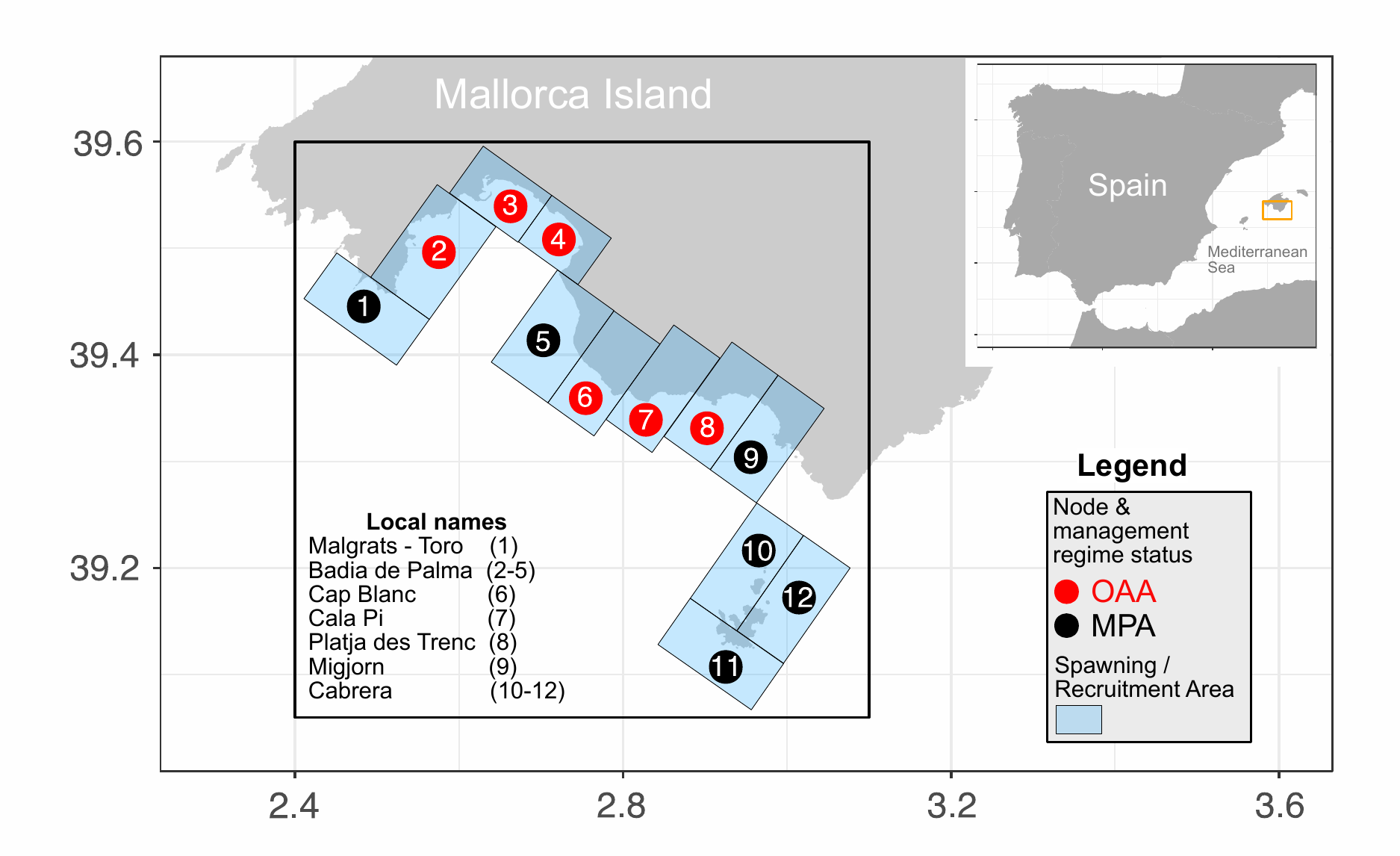}
	\caption{Map of the study area. The orange box in the top-right map shows the
		general location of the study area. The 12 nodes and management regime
		status are shown in the main map. Circles symbolising Marine Protected
		Areas (MPAs) and Open Access Areas (OAAs) are coloured in black and red,
		respectively. The areas shaded in blue show the zones established as
		spawning and recruiting areas. In these zones the number of exported
		larvae and the number of larvae recruited were summarised in model
		simulations.}
	\label{fig:fig1}
\end{figure}

Our case-study refers to the south-eastern coast of Mallorca
Island, in the Western Mediterranean Sea (Fig. \ref{fig:fig1}). In this region, we
identified 12 coastal sites (hereafter nodes) managed under two
different regimes: MPA and OAA (Table 1, Fig. \ref{fig:fig1}). Five MPAs protect the
most important ecological assets of the island (e.g., well preserved
seagrass meadows, rocky reefs), including the National
Marine-Terrestrial Park of Cabrera (a fully no-take MPA for recreational
fisheries since 1991). For this study, two small and nearby MPAs
(Malgrats and Toro) have been considered as a single MPA. A complete
description of environmental and hydrodynamic characteristics of the
area can be found in Basterretxea et al. (2012).

\begin{tiny}
	\begin{longtable}[]{@{}llllllll@{}}	
		\caption{Table 1. Areas (nodes) included in this study, their geographic location
			and management regime type (Marine Protected Area MPA and Open Access
			Area OAA). The available habitat and extension of protected area in each
			node (sq. km) and the size and abundance of the painted comber (Serranus
			scriba) are indicated.}
		\label{table:table1}
		\endhead
		\toprule
		\begin{minipage}[t]{0.05\columnwidth}\raggedright
			\textbf{Node}
			
			\textbf{ID}\strut
		\end{minipage} & \begin{minipage}[t]{0.10\columnwidth}\raggedright
			\textbf{Node name}\strut
		\end{minipage} & \begin{minipage}[t]{0.10\columnwidth}\raggedright
			\textbf{Latitude}
			
			\textbf{Longitude}\strut
		\end{minipage} & \begin{minipage}[t]{0.05\columnwidth}\raggedright
			\textbf{Manag.}
			
			\textbf{regime}\strut
		\end{minipage} & \begin{minipage}[t]{0.10\columnwidth}\raggedright
			\textbf{Available habitat (sq. km)}\strut
		\end{minipage} & \begin{minipage}[t]{0.10\columnwidth}\raggedright
			\textbf{Protected area of sea (sq. km)}\strut
		\end{minipage} & \begin{minipage}[t]{0.15\columnwidth}\raggedright
			\textbf{Fish size\\
				mean -- sd}
			
			\textbf{N}\strut
		\end{minipage} & \begin{minipage}[t]{0.15\columnwidth}\raggedright
			\textbf{Fish abundance (ind. sq. km habitat)\\
				mean - sd{~}}\strut
		\end{minipage}\tabularnewline
		\toprule
		\begin{minipage}[t]{0.05\columnwidth}\raggedright
			1\strut
		\end{minipage} & \begin{minipage}[t]{0.10\columnwidth}\raggedright
			Malgrats - Toro\strut
		\end{minipage} & \begin{minipage}[t]{0.10\columnwidth}\raggedright
			39.445N
			
			2.468E\strut
		\end{minipage} & \begin{minipage}[t]{0.05\columnwidth}\raggedright
			MPA\strut
		\end{minipage} & \begin{minipage}[t]{0.10\columnwidth}\raggedright
			0.75\strut
		\end{minipage} & \begin{minipage}[t]{0.10\columnwidth}\raggedright
			2.25\strut
		\end{minipage} & \begin{minipage}[t]{0.15\columnwidth}\raggedright
			135 -- 29.73
			
			214\strut
		\end{minipage} & \begin{minipage}[t]{0.15\columnwidth}\raggedright
			4531.36 -- 2758.46\strut
		\end{minipage}\tabularnewline
		\begin{minipage}[t]{0.05\columnwidth}\raggedright
			2\strut
		\end{minipage} & \begin{minipage}[t]{0.10\columnwidth}\raggedright
			Badia de Palma\strut
		\end{minipage} & \begin{minipage}[t]{0.10\columnwidth}\raggedright
			39.495N
			
			2.574E\strut
		\end{minipage} & \begin{minipage}[t]{0.05\columnwidth}\raggedright
			OAA\strut
		\end{minipage} & \begin{minipage}[t]{0.10\columnwidth}\raggedright
			9.74\strut
		\end{minipage} & \begin{minipage}[t]{0.10\columnwidth}\raggedright
			\strut
		\end{minipage} & \begin{minipage}[t]{0.15\columnwidth}\raggedright
			124 -- 27.68
			
			467\strut
		\end{minipage} & \begin{minipage}[t]{0.15\columnwidth}\raggedright
			96651.25 -- 46475.54\strut
		\end{minipage}\tabularnewline
		\begin{minipage}[t]{0.05\columnwidth}\raggedright
			3\strut
		\end{minipage} & \begin{minipage}[t]{0.10\columnwidth}\raggedright
			Badia de Palma\strut
		\end{minipage} & \begin{minipage}[t]{0.10\columnwidth}\raggedright
			39.538N
			
			2.658E\strut
		\end{minipage} & \begin{minipage}[t]{0.05\columnwidth}\raggedright
			OAA\strut
		\end{minipage} & \begin{minipage}[t]{0.10\columnwidth}\raggedright
			8.08\strut
		\end{minipage} & \begin{minipage}[t]{0.10\columnwidth}\raggedright
			\strut
		\end{minipage} & \begin{minipage}[t]{0.15\columnwidth}\raggedright
			124 -- 30.64
			
			279\strut
		\end{minipage} & \begin{minipage}[t]{0.15\columnwidth}\raggedright
			68074.11 -- 16431.04\strut
		\end{minipage}\tabularnewline
		\begin{minipage}[t]{0.05\columnwidth}\raggedright
			4\strut
		\end{minipage} & \begin{minipage}[t]{0.10\columnwidth}\raggedright
			Badia de Palma\strut
		\end{minipage} & \begin{minipage}[t]{0.10\columnwidth}\raggedright
			39.502N
			
			2.720E\strut
		\end{minipage} & \begin{minipage}[t]{0.05\columnwidth}\raggedright
			OAA\strut
		\end{minipage} & \begin{minipage}[t]{0.10\columnwidth}\raggedright
			18.12\strut
		\end{minipage} & \begin{minipage}[t]{0.10\columnwidth}\raggedright
			\strut
		\end{minipage} & \begin{minipage}[t]{0.15\columnwidth}\raggedright
			121 -- 29.76
			
			607\strut
		\end{minipage} & \begin{minipage}[t]{0.15\columnwidth}\raggedright
			152634.46 -- 36841.35\strut
		\end{minipage}\tabularnewline
		\begin{minipage}[t]{0.05\columnwidth}\raggedright
			5\strut
		\end{minipage} & \begin{minipage}[t]{0.10\columnwidth}\raggedright
			Badia de Palma\strut
		\end{minipage} & \begin{minipage}[t]{0.10\columnwidth}\raggedright
			39.405N
			
			2.701E\strut
		\end{minipage} & \begin{minipage}[t]{0.05\columnwidth}\raggedright
			MPA\strut
		\end{minipage} & \begin{minipage}[t]{0.10\columnwidth}\raggedright
			9.24\strut
		\end{minipage} & \begin{minipage}[t]{0.10\columnwidth}\raggedright
			23.94\strut
		\end{minipage} & \begin{minipage}[t]{0.15\columnwidth}\raggedright
			144 -- 24.95
			
			135\strut
		\end{minipage} & \begin{minipage}[t]{0.15\columnwidth}\raggedright
			111311.57 -- 73869.36\strut
		\end{minipage}\tabularnewline
		\begin{minipage}[t]{0.05\columnwidth}\raggedright
			6\strut
		\end{minipage} & \begin{minipage}[t]{0.10\columnwidth}\raggedright
			Cap Blanc\strut
		\end{minipage} & \begin{minipage}[t]{0.10\columnwidth}\raggedright
			39.358N
			
			2.753E\strut
		\end{minipage} & \begin{minipage}[t]{0.05\columnwidth}\raggedright
			OAA\strut
		\end{minipage} & \begin{minipage}[t]{0.10\columnwidth}\raggedright
			1.32\strut
		\end{minipage} & \begin{minipage}[t]{0.10\columnwidth}\raggedright
			\strut
		\end{minipage} & \begin{minipage}[t]{0.15\columnwidth}\raggedright
			143 -- 26.91
			
			238\strut
		\end{minipage} & \begin{minipage}[t]{0.15\columnwidth}\raggedright
			8215.92 -- 5417.97\strut
		\end{minipage}\tabularnewline
		\begin{minipage}[t]{0.05\columnwidth}\raggedright
			7\strut
		\end{minipage} & \begin{minipage}[t]{0.10\columnwidth}\raggedright
			Cala Pi\strut
		\end{minipage} & \begin{minipage}[t]{0.10\columnwidth}\raggedright
			39.333N
			
			2.823E\strut
		\end{minipage} & \begin{minipage}[t]{0.05\columnwidth}\raggedright
			OAA\strut
		\end{minipage} & \begin{minipage}[t]{0.10\columnwidth}\raggedright
			5.76\strut
		\end{minipage} & \begin{minipage}[t]{0.10\columnwidth}\raggedright
			\strut
		\end{minipage} & \begin{minipage}[t]{0.15\columnwidth}\raggedright
			119 -- 25.56
			
			84\strut
		\end{minipage} & \begin{minipage}[t]{0.15\columnwidth}\raggedright
			64827.08 -- 26730.27\strut
		\end{minipage}\tabularnewline
		\begin{minipage}[t]{0.05\columnwidth}\raggedright
			8\strut
		\end{minipage} & \begin{minipage}[t]{0.10\columnwidth}\raggedright
			Platja des Trenc\strut
		\end{minipage} & \begin{minipage}[t]{0.10\columnwidth}\raggedright
			39.328N
			
			2.904E\strut
		\end{minipage} & \begin{minipage}[t]{0.05\columnwidth}\raggedright
			OAA\strut
		\end{minipage} & \begin{minipage}[t]{0.10\columnwidth}\raggedright
			14.69\strut
		\end{minipage} & \begin{minipage}[t]{0.10\columnwidth}\raggedright
			\strut
		\end{minipage} & \begin{minipage}[t]{0.15\columnwidth}\raggedright
			120 -- 24.77
			
			37\strut
		\end{minipage} & \begin{minipage}[t]{0.15\columnwidth}\raggedright
			127115.42 -- 100846.30\strut
		\end{minipage}\tabularnewline
		\begin{minipage}[t]{0.05\columnwidth}\raggedright
			9\strut
		\end{minipage} & \begin{minipage}[t]{0.10\columnwidth}\raggedright
			Migjorn\strut
		\end{minipage} & \begin{minipage}[t]{0.10\columnwidth}\raggedright
			39.308N
			
			2.966E\strut
		\end{minipage} & \begin{minipage}[t]{0.05\columnwidth}\raggedright
			MPA\strut
		\end{minipage} & \begin{minipage}[t]{0.10\columnwidth}\raggedright
			10.47\strut
		\end{minipage} & \begin{minipage}[t]{0.10\columnwidth}\raggedright
			223.32\strut
		\end{minipage} & \begin{minipage}[t]{0.15\columnwidth}\raggedright
			112 -27.36
			
			63\strut
		\end{minipage} & \begin{minipage}[t]{0.15\columnwidth}\raggedright
			96173.66 -- 46896.00\strut
		\end{minipage}\tabularnewline
		\begin{minipage}[t]{0.05\columnwidth}\raggedright
			10\strut
		\end{minipage} & \begin{minipage}[t]{0.10\columnwidth}\raggedright
			Cabrera\strut
		\end{minipage} & \begin{minipage}[t]{0.10\columnwidth}\raggedright
			39.198N
			
			2.945E\strut
		\end{minipage} & \begin{minipage}[t]{0.05\columnwidth}\raggedright
			MPA\strut
		\end{minipage} & \begin{minipage}[t]{0.10\columnwidth}\raggedright
			4.51\strut
		\end{minipage} & \begin{minipage}[t]{0.10\columnwidth}\raggedright
			\strut
		\end{minipage} & \begin{minipage}[t]{0.15\columnwidth}\raggedright
			151 -27.72
			
			354\strut
		\end{minipage} & \begin{minipage}[t]{0.15\columnwidth}\raggedright
			71691.32 -- 17304.12\strut
		\end{minipage}\tabularnewline
		\begin{minipage}[t]{0.05\columnwidth}\raggedright
			11\strut
		\end{minipage} & \begin{minipage}[t]{0.10\columnwidth}\raggedright
			Cabrera\strut
		\end{minipage} & \begin{minipage}[t]{0.10\columnwidth}\raggedright
			39.116N
			
			2.914E\strut
		\end{minipage} & \begin{minipage}[t]{0.05\columnwidth}\raggedright
			MPA\strut
		\end{minipage} & \begin{minipage}[t]{0.10\columnwidth}\raggedright
			0.98\strut
		\end{minipage} & \begin{minipage}[t]{0.10\columnwidth}\raggedright
			86.8\strut
		\end{minipage} & \begin{minipage}[t]{0.15\columnwidth}\raggedright
			164 -26.89
			
			28\strut
		\end{minipage} & \begin{minipage}[t]{0.15\columnwidth}\raggedright
			15518.97 -- 3745.81\strut
		\end{minipage}\tabularnewline
		\begin{minipage}[t]{0.05\columnwidth}\raggedright
			12\strut
		\end{minipage} & \begin{minipage}[t]{0.10\columnwidth}\raggedright
			Cabrera\strut
		\end{minipage} & \begin{minipage}[t]{0.10\columnwidth}\raggedright
			39.162N
			
			3.010E\strut
		\end{minipage} & \begin{minipage}[t]{0.05\columnwidth}\raggedright
			MPA\strut
		\end{minipage} & \begin{minipage}[t]{0.10\columnwidth}\raggedright
			3.3\strut
		\end{minipage} & \begin{minipage}[t]{0.10\columnwidth}\raggedright
			\strut
		\end{minipage} & \begin{minipage}[t]{0.15\columnwidth}\raggedright
			163 -24.42
			
			14\strut
		\end{minipage} & \begin{minipage}[t]{0.15\columnwidth}\raggedright
			52523.67 -- 12677.63\strut
		\end{minipage}\tabularnewline
		\bottomrule
	\end{longtable}
\end{tiny}

\subsection{Hydrodynamic model}
The coupled hydrodynamic-biological IBM used in this work has been
previously described and validated {(Basterretxea et al. 2012)}. Some
outstanding features of the modelling approach include the simulation of
the hydrodynamics around Mallorca Island using a three-dimensional
density-resolving model, based on the Princeton Ocean Model (POM), with
a resolution of 200 m in horizontal and 25 sigma layers in vertical from
2000 to 2009. The model was executed offline and included wind forcing
using 12-hour maps for the same period.{~}

\subsection{Potential larval exchange matrix}

A total 19 325 passive tracers were released at weekly intervals from
200 m equispaced grid points located over seagrass ⁄ rocky bottoms
(\textless40 m depth) for a 6-month period (March-- August) each year.
The depth of release of the particles was random within the first 10 m
of the water column. Particles were treated as passive, neutrally
buoyant, flowing near the surface being freely transported by the
currents for 21 days. A reflective boundary condition was used to
prevent particles from moving onto land. At the end of this period the
simulation was stopped. The final position (latitude, longitude and
depth) of the passive tracers was used to determine their fate. Larvae
within the 12 previously defined areas were assumed to successfully
settle. A potential larval matrix was obtained representing the
probability of connection between the 12 sites. This matrix can be
associated to an ensemble of coastal fish species inhabiting the rocky,
sand and seagrass meadows of the Mediterranean Sea; with summer spawning
preferences; and a PLD of 21-days (Table S1).{~}

\subsection{Realised larval exchange matrix}

The painted comber \emph{Serranus scriba} was selected as model species
for calculating the realised larval exchange matrix in this modelling
approach because its breeding season spans from May to August in the
Balearic Islands {(Alonso-Fernández et al. 2011). Moreover, }its
dispersal is restricted to early-life stages {(March et al. 2010)} that
are pelagic during \emph{ca.} three weeks {(Macpherson \& Raventós
2006).} Here, we have developed a total egg production model based on
the individual, which incorporates valuable information of basic
reproductive parameters of \emph{S. scriba}, like size- and
site-dependent fecundity or a long spawning season, as revealed by the
empirical data {(Alonso-Fernández et al. 2011, Alós et al. 2013, 2014)}.
Potential fecundity (oocytes/ female), density (individuals
m\textsuperscript{-2}), proportion of females and proportion of mature
females were used as proxies in the model to assess the total egg
abundance (See Supplementary material for a detailed model explanation).
The model output was used to calculate the realised larval exchange
matrix by weighting the potential larval matrix with the spatial
variability in egg abundance (oocytes m\textsuperscript{-2}). In
consequence, the egg abundance is a function of adult population
abundance and reproductive characteristics, that are a function of the
extent of available habitat (rocky and seagrass meadows down to 40 m
depth) and the level of protection (MPA \emph{vs.} OAA, see Table 1).
Since information on vertical movements or the orientation of the larvae
towards sound or light stimuli for the painted comber is non-existent,
directed horizontal movement and the vertical movement of the larvae
were not parameterised in the hydrodynamic simulations.

\subsection{Matrices, graphs and networks as representations of connected systems}

Any adjacency matrix can be represented as a graph with its nodes and
edges and, consequently, they are equivalent representations of a
connected system. Hereafter, we will refer to connected larval flow
systems as larval connectivity networks. The potential larval exchange
matrix, and its correspondent graph, represents the larval connectivity
network of an ensemble of species with a common PLD (Table S1). The
realised larval exchange matrix, and its correspondent graph, represents
the larval connectivity network of the painted comber, as a model
species. Although the visual representation of adjacency matrices on
their own provides valuable information about the links within the
entire network (i.e., probabilities or absolute numbers), the use of
graph theory provides key information about the role of each node within
a larval connectivity network (Fig S1, Minor and Urban 2008). To explore
the potential of graph theory methods, we calculated centrality measures
to understand the effect of the distinct levels of connectivity derived
from potential and realised larval connectivity networks.{~}

\subsection{Road map for the selection of retention indices and centrality measures}

A road map was established for the selection of retention indices and
centrality measures based on their usefulness and importance for
conservation, recreational fisheries or management. This roadmap is
summarized as follows (Fig \ref{fig:fig2}):

\begin{figure}
	\centering
	\includegraphics[width=1\linewidth]{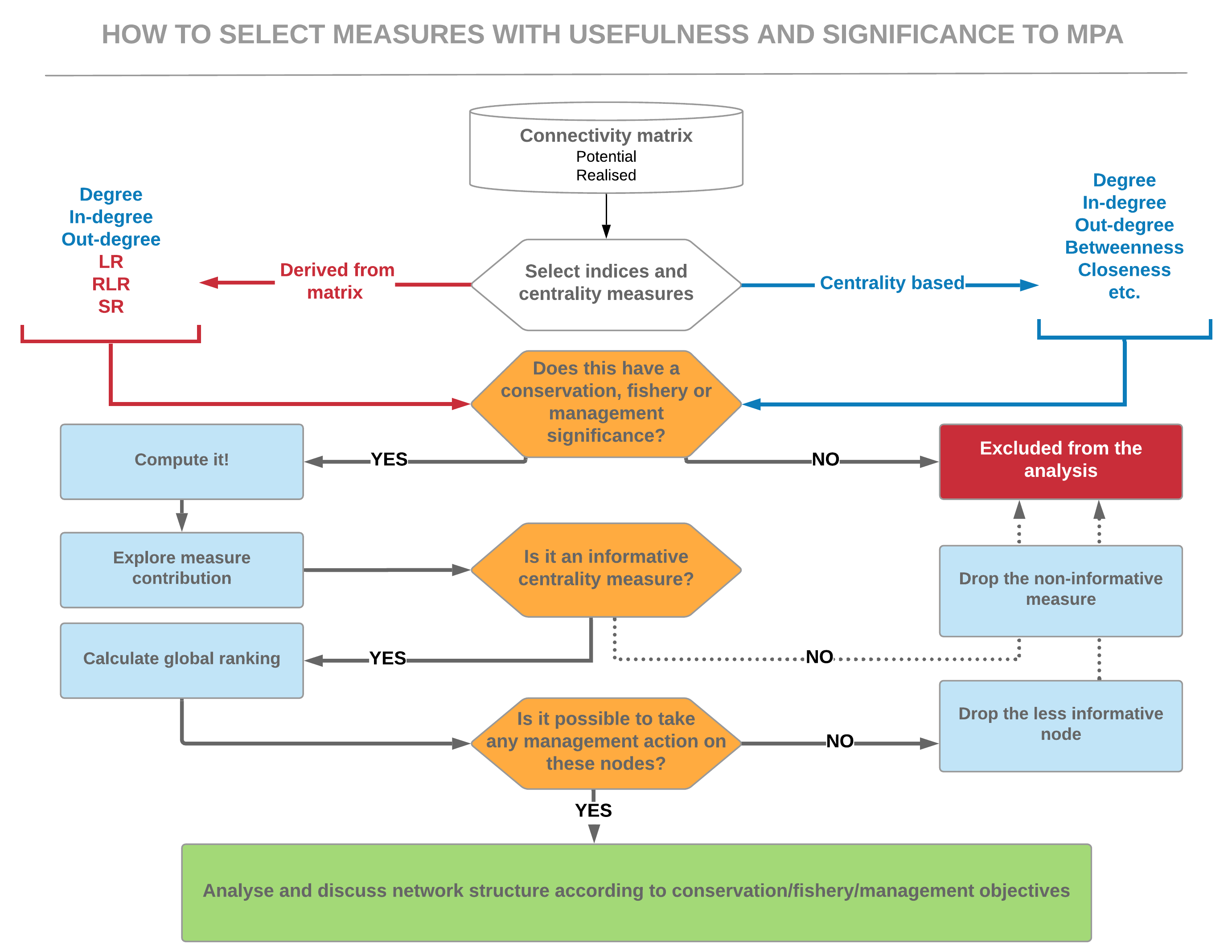}
	\caption{Diagram synthesising the selection criteria for retention indices and
		centrality measures based on usefulness and significance for
		conservation, fishery or management.}
	\label{fig:fig2}
\end{figure}

\begin{enumerate}
	\item Obtain the larval connectivity network (e.g. by species, functional
	groups, etc.).
	
	\item Select the retention indices and centrality measures that can be
	calculated according to the characteristics of each connectivity network
	(e.g., strength can only be calculated in weighted networks, or
	closeness cannot be calculated in disconnected networks).
	
	\item Calculate indices and measures that may be relevant to conservation,
	recreational fisheries or management (see rationale column in Tables 2 y
	3).
	
	\item Exploratory data analysis to assist decisions on which indices and
	measures are informative (e.g., summary statistics, correlation
	analysis, etc.).
	
	\item The indices and measures are normalised and the nodes (locations) of
	the network are ranked globally.
	
	\item Management decisions could be made, and management scenarios be
	designed according to the overall ranking of the nodes.
	
	\item The structure of the resulting connectivity network is analyzed and
	discussed according to the conservation, recreational fisheries or
	management objectives.
\end{enumerate}

	\begin{tiny}
	\begin{longtable}[]{@{}llll@{}}
		\caption{Self-persistence indices of local marine populations (as
			presented by Lett et al. 2015)}
		\label{table:table2}
		\endhead
		\toprule
		\begin{minipage}[t]{0.22\columnwidth}\raggedright
			\textbf{Index}\strut
		\end{minipage} & \begin{minipage}[t]{0.22\columnwidth}\raggedright
			\textbf{Definition}\strut
		\end{minipage} & \begin{minipage}[t]{0.22\columnwidth}\raggedright
			\textbf{Calculation}\strut
		\end{minipage} & \begin{minipage}[t]{0.22\columnwidth}\raggedright
			\textbf{Rationale}\strut
		\end{minipage}\tabularnewline
		\toprule
		\begin{minipage}[t]{0.22\columnwidth}\raggedright
			Local Retention (LR)\strut
		\end{minipage} & \begin{minipage}[t]{0.22\columnwidth}\raggedright
			The ratio of locally produced settlement to local egg production.\strut
		\end{minipage} & \begin{minipage}[t]{0.22\columnwidth}\raggedright
			LR is equal to the diagonal elements of the connectivity matrix.\strut
		\end{minipage} & \begin{minipage}[t]{0.22\columnwidth}\raggedright
			LR does not depend on egg production in the different habitat patches,
			meaning it is independent of temporal changes in adult population
			size.\strut
		\end{minipage}\tabularnewline
		\begin{minipage}[t]{0.22\columnwidth}\raggedright
			Relative Local Retention (RLR)\strut
		\end{minipage} & \begin{minipage}[t]{0.22\columnwidth}\raggedright
			The ratio of locally produced settlement to all settlement of local
			origin.\strut
		\end{minipage} & \begin{minipage}[t]{0.22\columnwidth}\raggedright
			RLR is equal to the diagonal elements of the connectivity matrix divided
			by the sum of the corresponding column of the connectivity matrix.\strut
		\end{minipage} & \begin{minipage}[t]{0.22\columnwidth}\raggedright
			RLR does not depend on egg production, but it does depend on the
			connectivity between the focal patch and the ensemble of patches in the
			system.\strut
		\end{minipage}\tabularnewline
		\begin{minipage}[t]{0.22\columnwidth}\raggedright
			Self-Recruitment (SR)\strut
		\end{minipage} & \begin{minipage}[t]{0.22\columnwidth}\raggedright
			The ratio of locally produced settlement to settlement of all origins at
			a site.\strut
		\end{minipage} & \begin{minipage}[t]{0.22\columnwidth}\raggedright
			If egg production is uniform over sites, then SR is equal to the
			diagonal elements of the connectivity matrix divided by the sum of the
			corresponding row of the connectivity matrix. If not, then the elements
			of the dispersal matrix must be weighted by the number of eggs
			produced.\strut
		\end{minipage} & \begin{minipage}[t]{0.22\columnwidth}\raggedright
			SR depends on egg production in the different habitat patches (relative
			to that of the focal patch) and therefore on temporal changes in
			population size. SR also depends on connectivity between every patch and
			the focal patch.\strut
		\end{minipage}\tabularnewline
		\bottomrule
	\end{longtable}
\end{tiny}

\subsection{Retention indices and centrality measures}

Following the roadmap (Fig. \ref{fig:fig2}), 3 retention indices and 9 centrality
measures were considered to have potential relevance to conservation,
recreational fisheries or management. To capture hydrodynamic retention
and, consequently, larval retention in our system, we calculated 3
different retention indices for the potential and realised larval
connectivity networks following Lett et al. (2015): local retention
(LR); relative local retention (RLR) and self-recruitment (SR).
Retention indices are defined in Table 2 according {Lett et al. (2015)}.

With the aim of highlighting the most important nodes for larval
connectivity within the network, we calculated 9 measures of centrality
for the potential and realised networks: In-strength; Out-strength;
Strength; Betweenness, Eigenvector centrality; Closeness; Kleinberg's
hub centrality score (hereafter Hub score); Kleinberg's authority
centrality score (hereafter Authority score); and Page Rank. These
centrality measures were selected as those potentially useful in larval
connectivity studies and come from a first selection that included all
existing measures of centrality, identified from a review of the
existing literature. A full description of each centrality measure
selected, their scope and ecological interpretation and references to
the original source, is provided in Table 3. Note that Out-strength and
In-strength correspond to the sum of all eggs released and larvae
recruited from each node, respectively, and Strength correspond to the
sum of In-strength and Out-strength. In consequence, these 3 centrality
measures can be calculated either from a graph or directly from an
adjacency matrix. Higher values in centrality measures indicate greater
"connectivity" (\emph{sensu lato}). Connectivity is not understood as
larval connectivity but as connectivity between the nodes of the
network. Therefore, each centrality measure informs of node connectivity
properties within the network and provides different ecological
interpretation of network connectivity. The selection and prioritisation
of one measure over the other should be carefully considered depending
on management and conservation local targets (Fig. \ref{fig:fig2}, Table 3).

\begin{tiny}
	\begin{longtable}[]{@{}llll@{}}
		\caption{Measures of centrality, definitions and rationales in a context
			of ecological - larval connectivity. See references for additional
			information. The formula used in each centrality measurement can be
			found in the documentation of the "igraph" package for R.}
		\label{table:table3}
		\endhead
		\toprule
		\begin{minipage}[t]{0.22\columnwidth}\raggedright
			\textbf{measure}\strut
		\end{minipage} & \begin{minipage}[t]{0.22\columnwidth}\raggedright
			\textbf{reference}\strut
		\end{minipage} & \begin{minipage}[t]{0.22\columnwidth}\raggedright
			\textbf{definition}\strut
		\end{minipage} & \begin{minipage}[t]{0.22\columnwidth}\raggedright
			\textbf{rationale}\strut
		\end{minipage}\tabularnewline
		\toprule
		\begin{minipage}[t]{0.22\columnwidth}\raggedright
			Strength\strut
		\end{minipage} & \begin{minipage}[t]{0.22\columnwidth}\raggedright
			Barrat et al. 2004\strut
		\end{minipage} & \begin{minipage}[t]{0.22\columnwidth}\raggedright
			Also named weighted degree. The node degree is the number of relations
			(edges) of the nodes. In weighted networks, node Strength is the sum of
			weights of links connected to the node.\strut
		\end{minipage} & \begin{minipage}[t]{0.22\columnwidth}\raggedright
			Strength indicates a node is involved in many important (by weight)
			interactions with other nodes. Nodes with high Strength can be acting as
			keystones since they are connected by egg production and recruitment to
			many neighbouring nodes.\strut
		\end{minipage}\tabularnewline
		\begin{minipage}[t]{0.22\columnwidth}\raggedright
			In-strength\strut
		\end{minipage} & \begin{minipage}[t]{0.22\columnwidth}\raggedright
			Barrat et al. 2004\strut
		\end{minipage} & \begin{minipage}[t]{0.22\columnwidth}\raggedright
			{~}In directed networks, the In-strength is the sum of inward link
			weights.\strut
		\end{minipage} & \begin{minipage}[t]{0.22\columnwidth}\raggedright
			Nodes with high In-strength can be acting as important nursery areas,
			sensitive recruitment or settlement zones.\strut
		\end{minipage}\tabularnewline
		\begin{minipage}[t]{0.22\columnwidth}\raggedright
			Out-strength\strut
		\end{minipage} & \begin{minipage}[t]{0.22\columnwidth}\raggedright
			Barrat et al. 2004\strut
		\end{minipage} & \begin{minipage}[t]{0.22\columnwidth}\raggedright
			{~}In directed networks, the Out-strength is the sum of outward link
			weights.\strut
		\end{minipage} & \begin{minipage}[t]{0.22\columnwidth}\raggedright
			Nodes with high Out-strength can be acting as essential spawning
			habitats. Genetically, sub-populations inhabiting nodes with high
			Out-strength have a high probability to spread genes to other
			sub-population in the network.\strut
		\end{minipage}\tabularnewline
		\begin{minipage}[t]{0.22\columnwidth}\raggedright
			{~}Closeness\strut
		\end{minipage} & \begin{minipage}[t]{0.22\columnwidth}\raggedright
			Freeman 1979\strut
		\end{minipage} & \begin{minipage}[t]{0.22\columnwidth}\raggedright
			Closeness centrality indicates how long it will take for information
			from a given node to reach other nodes in the network.\strut
		\end{minipage} & \begin{minipage}[t]{0.22\columnwidth}\raggedright
			Nodes with a higher Closeness have a high probability of exporting
			propagules to their nearest neighbouring nodes.\strut
		\end{minipage}\tabularnewline
		\begin{minipage}[t]{0.22\columnwidth}\raggedright
			~Betweenness\strut
		\end{minipage} & \begin{minipage}[t]{0.22\columnwidth}\raggedright
			Freeman 1979\strut
		\end{minipage} & \begin{minipage}[t]{0.22\columnwidth}\raggedright
			Betweenness centrality is a measure of the influence of a node over the
			flow of information between every pair of nodes under the assumption
			that information primarily flows over the shortest paths between
			them.\strut
		\end{minipage} & \begin{minipage}[t]{0.22\columnwidth}\raggedright
			While high release of propagules and high recruitment of larvae is
			important, it's not everything. Nodes with high Betweenness centralities
			have been termed "bottlenecks" or "bridges" and they are preventing the
			fragmentation of the network. A node acting as a bridge between two
			well-differentiated subpopulations should have a high Betweenness.\strut
		\end{minipage}\tabularnewline
		\begin{minipage}[t]{0.22\columnwidth}\raggedright
			Eigenvector centrality\strut
		\end{minipage} & \begin{minipage}[t]{0.22\columnwidth}\raggedright
			Bonacich 1987\strut
		\end{minipage} & \begin{minipage}[t]{0.22\columnwidth}\raggedright
			The Eigenvector centrality network metric takes into consideration not
			only how many connections a node has (i.e., its Degree or Strength), but
			also the centrality of the vertices that it is connected to.\strut
		\end{minipage} & \begin{minipage}[t]{0.22\columnwidth}\raggedright
			It is a measure of the influence of a node in a network. In general, a
			connection to a well-connected node is more important than a connection
			to a poor connected node. Nodes with high Eigenvector centralities mean
			high-productivity and high-recruitment nodes connected to other
			high-productivity, high-recruitment nodes.{~}\strut
		\end{minipage}\tabularnewline
		\begin{minipage}[t]{0.22\columnwidth}\raggedright
			Kleinberg's hub centrality score (aka. Hub score)\strut
		\end{minipage} & \begin{minipage}[t]{0.22\columnwidth}\raggedright
			Kleinberg 2000\strut
		\end{minipage} & \begin{minipage}[t]{0.22\columnwidth}\raggedright
			The Hub score of a node shows how many highly informative nodes or
			authoritative nodes this node is pointing to.\strut
		\end{minipage} & \begin{minipage}[t]{0.22\columnwidth}\raggedright
			Nodes with high Hub scores mean nodes that recruit larvae from many
			other nodes in the network. Nodes that act as a nursery area very
			well-connected to other areas in the network.\strut
		\end{minipage}\tabularnewline
		\begin{minipage}[t]{0.22\columnwidth}\raggedright
			Kleinberg's authority centrality score (aka. Authority score)\strut
		\end{minipage} & \begin{minipage}[t]{0.22\columnwidth}\raggedright
			Kleinberg 2000\strut
		\end{minipage} & \begin{minipage}[t]{0.22\columnwidth}\raggedright
			The Authority score of a node is a measure of the amount of valuable
			information that this node holds.\strut
		\end{minipage} & \begin{minipage}[t]{0.22\columnwidth}\raggedright
			Nodes with high Authority scores mean nodes that export successful
			propagules to many other nodes in the network. Nodes that act as a
			well-connected source of propagules to other nodes in the network.\strut
		\end{minipage}\tabularnewline
		\begin{minipage}[t]{0.22\columnwidth}\raggedright
			Page Rank\strut
		\end{minipage} & \begin{minipage}[t]{0.22\columnwidth}\raggedright
			Brin and Page 1998\strut
		\end{minipage} & \begin{minipage}[t]{0.22\columnwidth}\raggedright
			Algorithm developed by Larry Page~and~Sergey Brin, founders of Google.
			Page Rank works by assigning importance to a webpage (node) if important
			pages (other nodes) point to it.\strut
		\end{minipage} & \begin{minipage}[t]{0.22\columnwidth}\raggedright
			It is interpreted similarly to the hub score. The approximate estimation
			of the importance of a node is based on the number and quality (weight)
			of the links pointing to it. The most important nodes are likely to
			recruit more larvae from other nodes in the network.\strut
		\end{minipage}\tabularnewline
		\bottomrule
	\end{longtable}
\end{tiny}

To facilitate the interpretation of the ranking of the nodes in a
network, firstly, each index or centrality measurement was normalised
between 0 and 1, with 1 corresponding to the highest connectivity value,
and secondly, the nodes were sorted in a descending order, with those
with the highest connectivity measures at the top. We also calculated
edge Betweenness centrality, a measure of arc centrality, defined as the
number of the shortest paths that go through an edge in a graph or
network {(Girvan \& Newman 2002)}. An arc with a high edge Betweenness
centrality score represents a bridging connector between two parts of a
network, the removal of which can affect communication between many
pairs of nodes through the shorter paths between them.

Following the roadmap described in section 2.5 (Fig. \ref{fig:fig2}), indices and
centrality measures were used to design a scenario where some
open-access could be converted into no-take areas and vice versa. The
biological parameters observed at Cabrera MPA (nodes 10 -- 12) were used
to scale-up egg production in the open-access areas to be converted to
no-take. To convert no-take into ``open to fisheries'' areas, the
average biological parameters of the open-access areas were used. The
characteristics and structure of the network were explored and the
impact of protection on larval network connectivity was discussed based
on the decisions made.

\subsection{Community detection}

In graph theory, it is possible to identify groups of nodes (i.e.,
modules) that probably share common properties and/or play similar roles
within the network. In networks where the distribution of links is
globally and locally inhomogeneous, some nodes have higher concentration
of links within special groups of nodes and low concentrations between
these groups. Here, we have considered the approach to community
detection using edge weights {(i.e., connection probability or larval
number, Brandes et al. 2008)}. When two or more network graphs are
compared, the weight differences among pairs result in differences in
cluster formation. Therefore, we quantified network modularity; the
existence of groups of locations that are highly interconnected among
themselves by proximity of links, and that are poorly connected to
locations in other modules {(Guimera \& Nunes Amaral 2005, Reichardt \&
Bornholdt 2006)}. Then, we calculated the optimal community structure
for the graph, in terms of maximal modularity score, to identify
clustering organization and identify communities {(following Brandes et
al. 2008)}. The number of communities and membership of each node were
used to visually display networks structure.{~}

All analyses were performed using the R language and environment for
statistical computing version 3.6.0, released 2019-04-26 {(R Core Team
2019). }Matrix, network graph and community detection analyses were
performed using R packages ``igraph'' v.1.2.4.1 and ``ConnMatTools''
v.0.3.3 {(Csardi \& Nepusz 2006, Kaplan et al. 2017). Network
visualisations were made with R packages: ``mapdata'' }v.2.3.0;
``ggplot2'' v.3.2.1 and ``ggraph'' v.2.0.0 {(Chang 2012, Wickham
2016).{~}}

\section{Results}

\subsection{Retention and self-recruitment}

\begin{figure}
	\centering
	\includegraphics[width=1\linewidth]{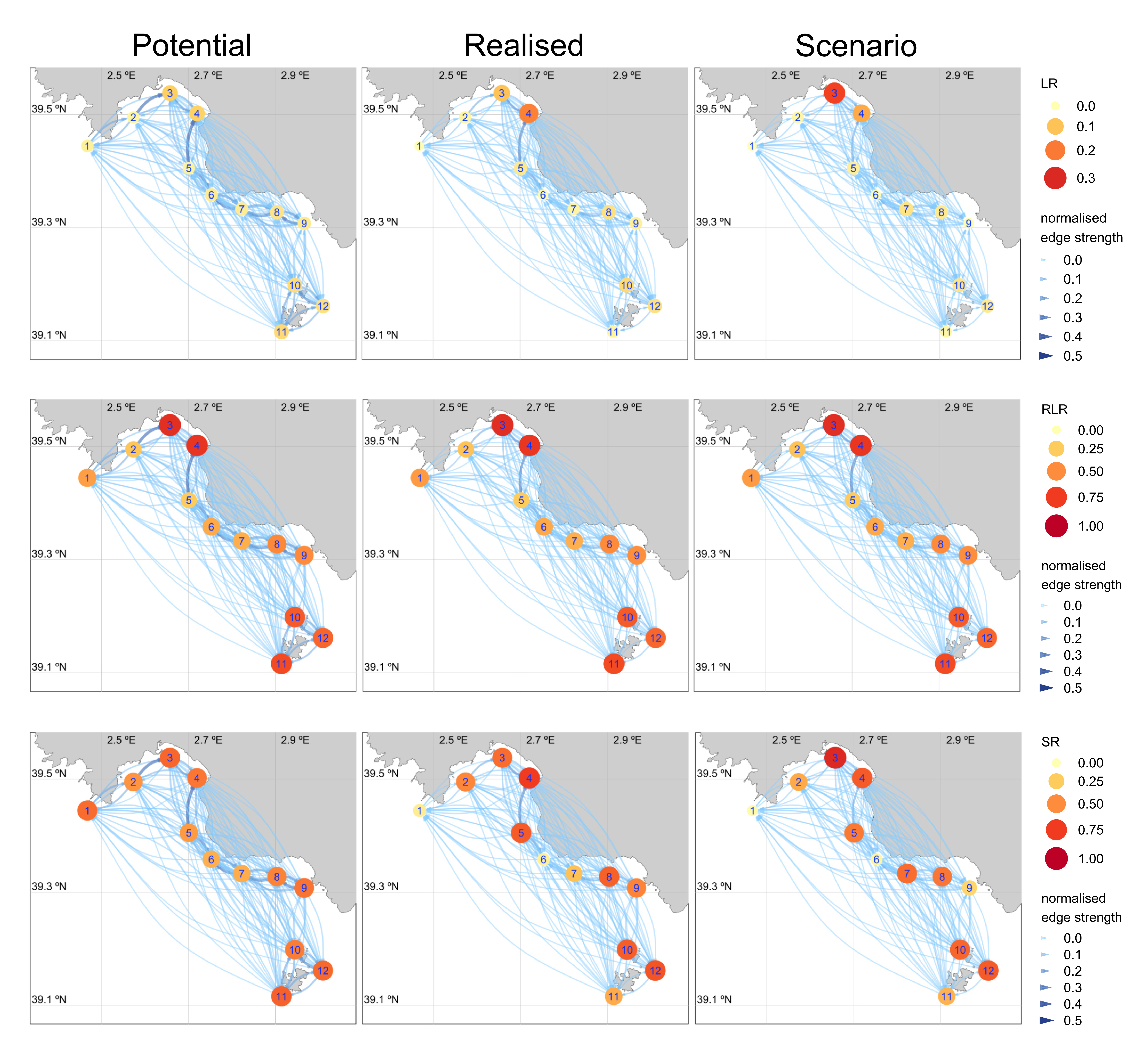}
	\caption{Local Recruitment (LR, top panel), Relative Local Recruitment (RLR,
		middle panel) and Self Recruitment (SR, bottom panel) indices for the
		potential, realised and scenario larval connectivity networks. Node size
		and colour symbolise index and edge width and colour symbolise
		normalised edge strength.}
	\label{fig:fig3}
\end{figure}

The nodes ranked according to their relative retention indices for
potential and realised larval connectivity and for the scenario networks
are shown in Tables S2, S3 andS4. In the case of the LR index, where the
sorting is based on absolute numbers and not on proportions, a slightly
different ranking of the nodes in the potential and realised networks is
observed, however nodes 3 and 4 are positioned as the most locally
retentive in both networks (Fig. \ref{fig:fig3}). These results are because nodes 3
and 4 are in the zone with the highest hydrodynamic retention and where
egg production is also the highest for any node in the network. LR is
sensitive to local patch (node) gamete production, but not to gamete
production from all metapopulations (the whole network). Therefore, LR
only provides information about local processes (i.e., node
processes).{~}

Nodes 3 and 4, located in the inner Palma Bay, show the highest RLR in
both the potential and realised larval connectivity networks. Indeed,
the same RLR node rank is obtained for these two networks since this
index is relative to reproductive output of the whole network, and the
results do not differ if either probabilities or measured demographic
weights based of the species are used (Fig. \ref{fig:fig3}).

SR, unlike RLR, is an index relative to the number of larvae arriving
and not to the gamete production of each node. In a system where egg
production is not uniform, and initial density is known, SR provides
valuable information on the relative importance of each node in the
larval connectivity system. The main SR differences between the
potential and the realised networks are occurring in nodes 5, 10 and 12
that are ranked higher in the realised network whereas node 3 descends
in the rank (Fig. \ref{fig:fig3}).{~}

\subsection{Analysis of centrality measures}

Out-strength, In-strength and Strength (the sum of the first two) are
measures of centrality with a marked influence on larval retention. The
nodes highlighted as important for these centrality measures are those
with the highest values of self-connection (i.e., LR) in the potential
and realised networks (Figs. \ref{fig:fig3} and S2).

\subsection{ Potential larval connectivity network}

In the case of the potential larval connectivity network, nodes 3 and 4
were the most important nodes when considering the Out-strength, i.e.,
the release of propagules that reach some recruitment node. If
self-loops (i.e., self-connections, LR) are not considered, the nodes
with the highest probability of emitting successful propagules were 5
and 7. Nodes 3 and 4 were also highlighted as those with the highest
probability of being successful nursery areas (Table S2, In-strength
column and Fig. S2).{~}

\begin{figure}
	\centering
	\includegraphics[width=1\linewidth]{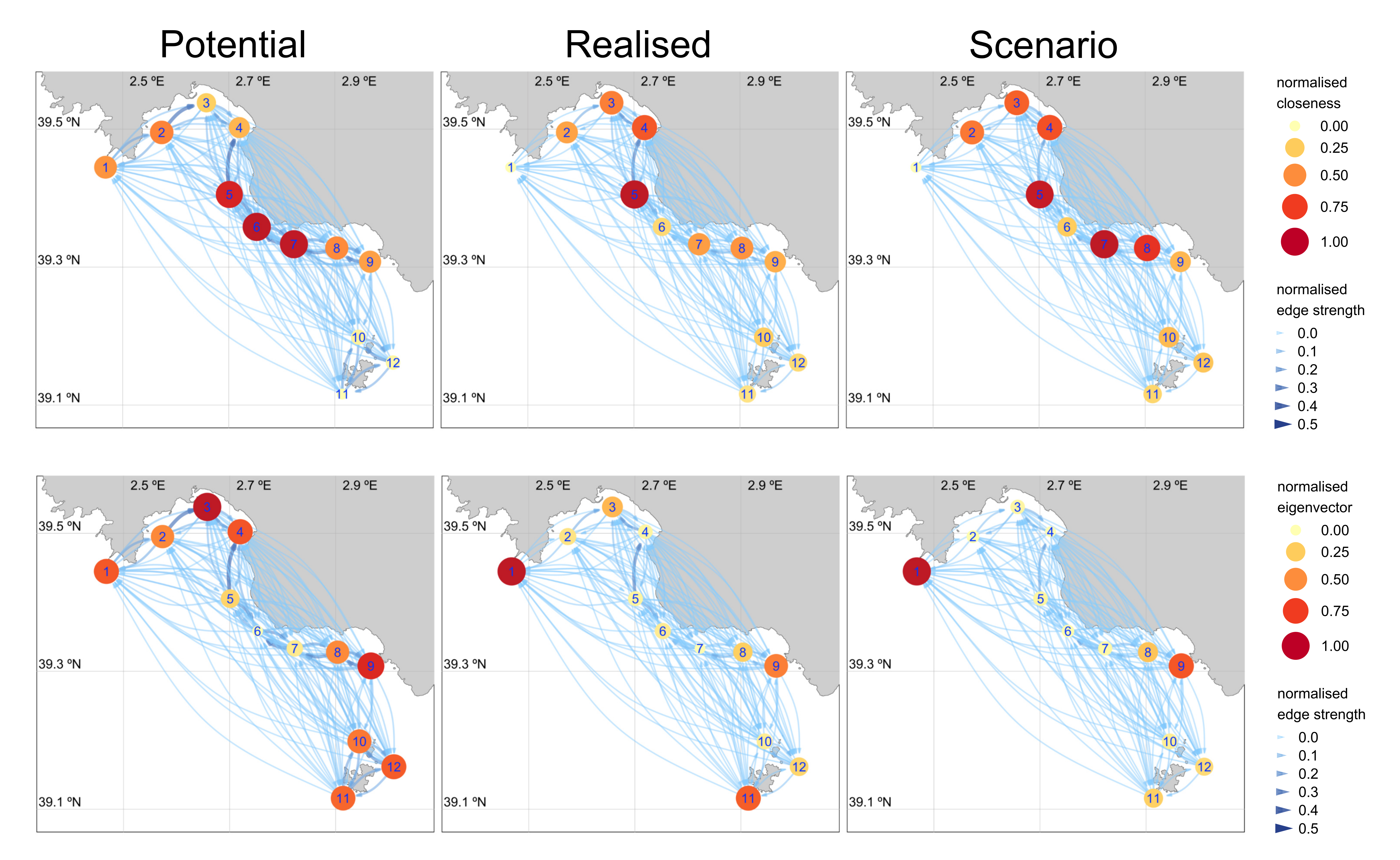}
	\caption{Closeness and Eigenvector centrality measures for the potential,
		realised and scenario larval connectivity networks. Node size and colour
		symbolise the centrality measure and edge width and colour symbolise
		normalised edge strength.}
	\label{fig:fig4}
\end{figure}

\begin{figure}
	\centering
	\includegraphics[width=1\linewidth]{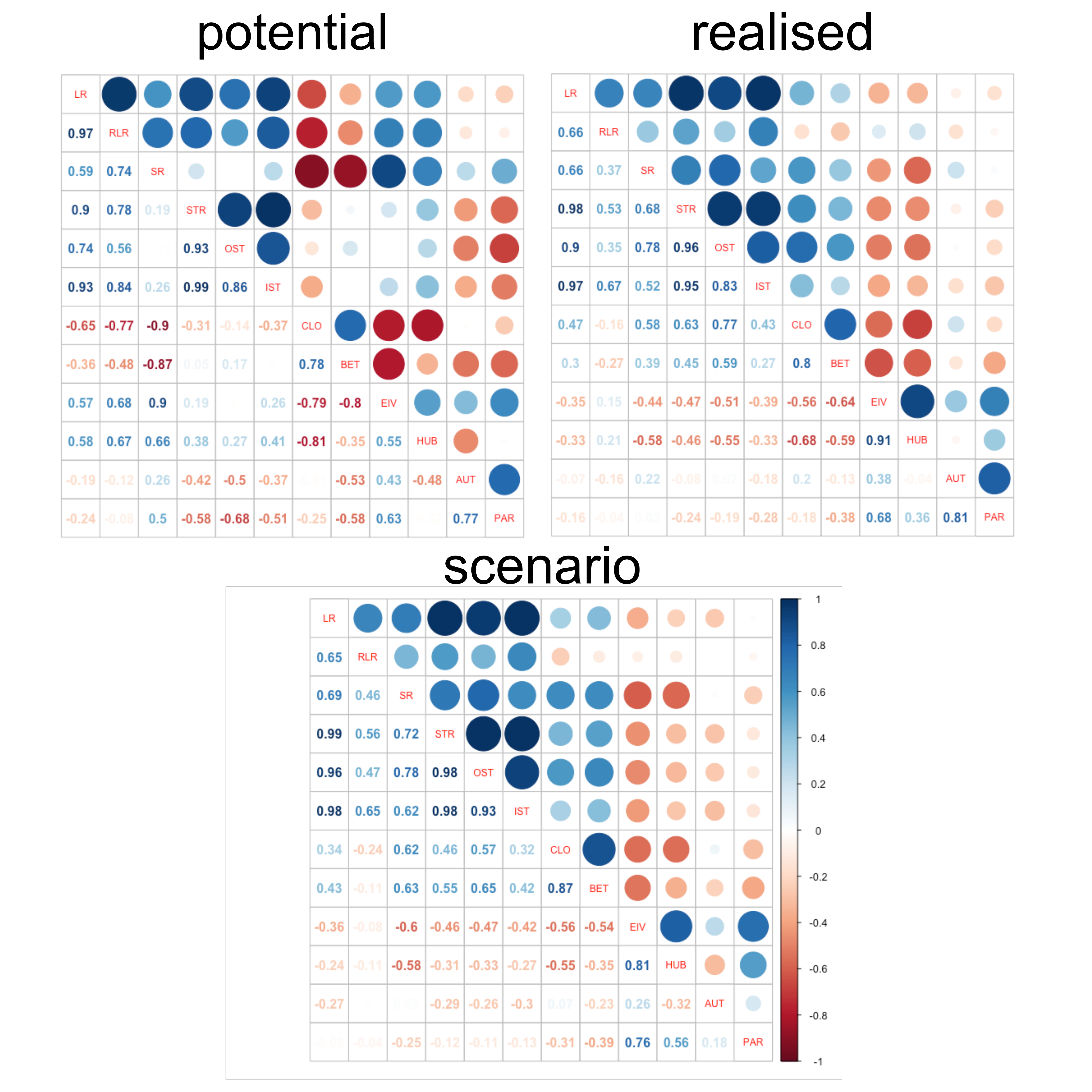}
	\caption{Heatmap of pairwise correlation values of 3 retention indices and 9
		centrality measures for the potential, realised and scenario larval
		connectivity networks. Local Retention (LR), Relative Local Retention
		(RLR), Self-Recruitment (SR), Strength (STR), Out-Strength (OST),
		In-Strength (IST), Closeness (CLO), Betweenness (BET), Eigenvector
		centrality (EIV), Hub score (HUB), Authority score (AUT) and Page Rank
		(PAR).}
	\label{fig:fig5}
\end{figure}

\begin{figure}
	\centering
	\includegraphics[width=1\linewidth]{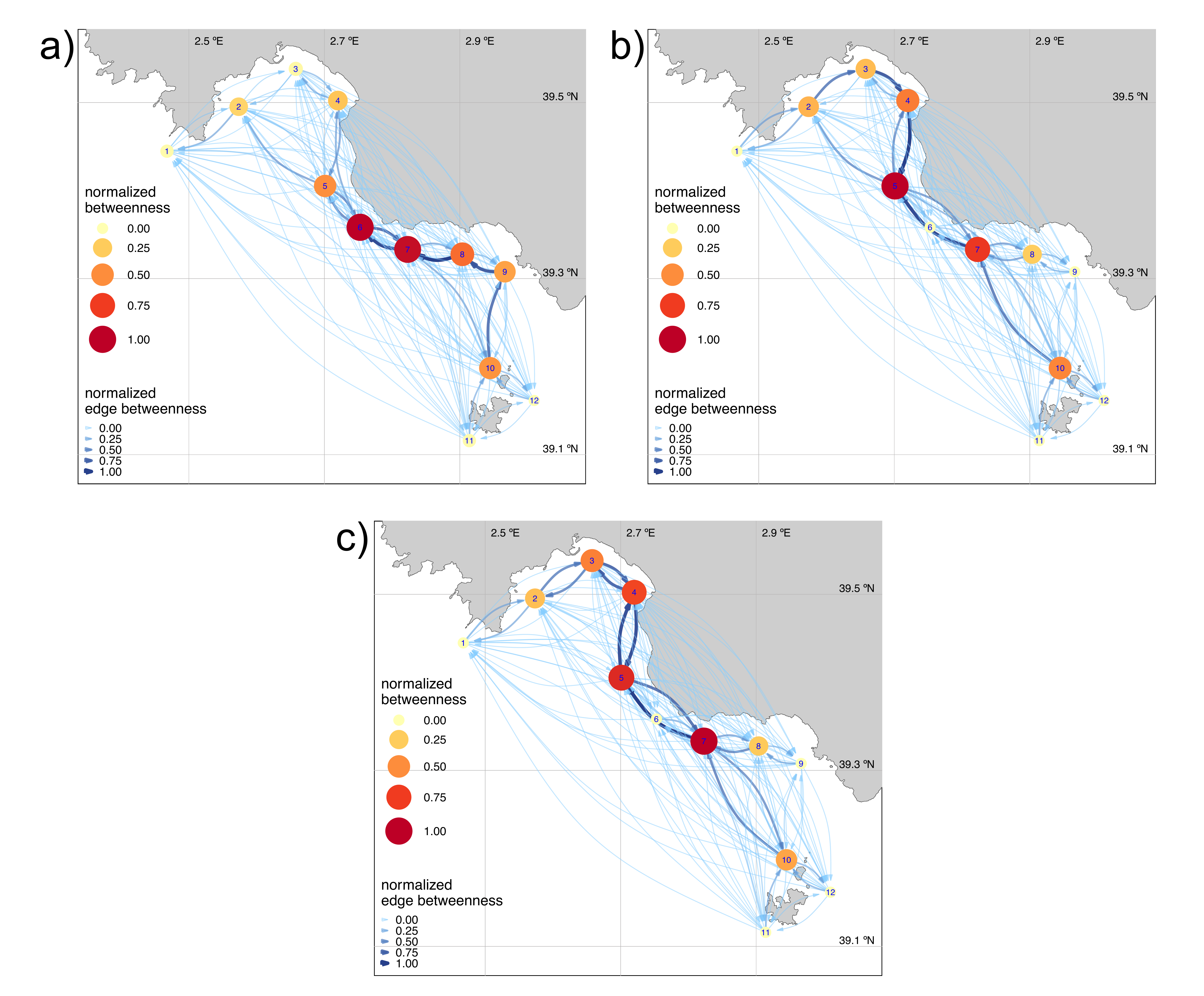}
	\caption{Betweenness centrality in the potential (a), realised (b) and scenario (c) larval connectivity networks. The circle size and colour indicate the importance of each node in
		maintaining network connectivity (the higher the value, the more likely
		a node is to act as a bridge between two sub-populations). The edge
		width represents how strong the connection between two nodes is. In the scenario network (c), open-access nodes 3 and 7 were converted to no-take (MPA) and no-take
		nodes 1 and 9 were ``open to fisheries'' (OAA).}
	\label{fig:fig6}
\end{figure}

In a network, a high Closeness value indicates a higher probability of
information propagation from one node to all other nodes in the network.
In a larval connectivity network, this probability will be higher for
those nodes that occupy areas of the network where hydrodynamics
facilitates larval transport to all other nodes. Accordingly, in the
potential larval connectivity network the nodes with highest Closeness
were 6 and 7, followed by 5 (Fig. \ref{fig:fig4}). There is a moderately high
correlation between the Closeness and Betweenness measures (r , Fig. \ref{fig:fig5}),
and nodes 6 and 7 also exhibited the highest Betweenness indicating
their importance as ``bridges'' between the nodes located north and
south of their location{~ }(Fig. \ref{fig:fig6} and Table S2). The connections
between the pair of nodes , , ,{~ }and{~ }exhibited the highest edge
Betweenness within the network. Excluding the self-loops, the strongest
connections were , ,{~ }and . These routes are the ones that concentrate
the largest flow of larvae between the different nodes of the potential
network. There is a strong inverse correlation between the Closeness and
SR index (r , Fig. \ref{fig:fig5}), which suggests that Closeness is a useful measure
of centrality since it allows the identification of nodes that
facilitate global connectivity in networks with a strong larval
retention influence.{~}

\begin{figure}
	\centering
	\includegraphics[width=1\linewidth]{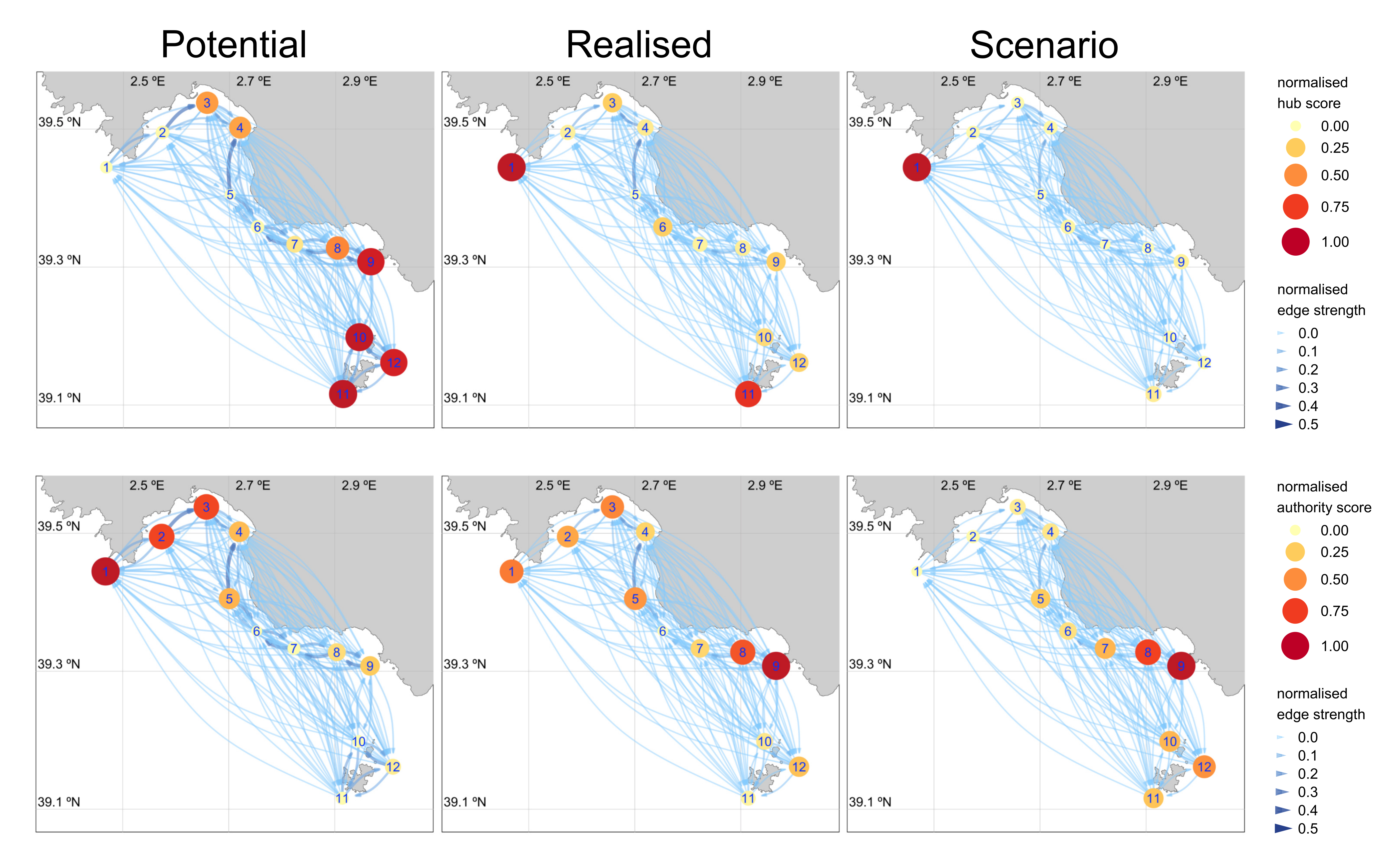}
	\caption{Kleinberg's Hub and Kleinberg's Authority centrality scores for the
		potential, realised and scenario larval connectivity networks. Node size
		and colour symbolise the centrality score and edge width and colour
		symbolise normalised edge strength.}
	\label{fig:fig7}
\end{figure}

Continuing with the potential larval connectivity network, nodes with
the highest Eigenvector, revealing the influence of a node in the
network, were 3 and 9, which also had high SR values (Fig. \ref{fig:fig3}). Indeed,
there is a strong correlation between these two metrics (r , Fig. \ref{fig:fig5}).
Nodes 10, 11 and 12 had the highest Hub-score, while nodes 1, 2 and 3
had the highest Authority-score (Fig. \ref{fig:fig7}). Nodes 1, 2 and 3 were also
highlighted by Page Rank (Fig. S3). These four measures (Eigenvector,
Hub and Authority score and Page Rank) could be useful to identify
critical nodes in a connectivity network. For example, node 1, with low
release and larval recruitment values, had a high Authority-score and
Page Rank (Table S2). This suggests that although the weight of its
connections to other nodes is low, these other nodes have the highest
connections in the network.

\subsection{  Realised larval connectivity network}

In the case of the realised larval connectivity network, node 4 held the
highest Strength and In-strength coinciding with what was observed in
the potential network (Tables S2 and S3 and Fig. S2). The node with the
highest probability of emitting successful larvae was node 4 (5 if
self-loops are not considered). According to Closeness, the
incorporation of demographic information shifts the importance of nodes
to those where more eggs are produced. Nodes 5, 4 and 3 had the highest
Closeness, i.e., the highest potential for connecting their larvae to
all the other nodes in the network (Fig. \ref{fig:fig4}). Betweenness was partly
correlated with Closeness centrality (r ). Regarding Betweenness, the
most important nodes in this network were 5 and 7 (Fig. \ref{fig:fig6}).
Surprisingly, the node with the highest Betweenness in the potential
network (node 6) had the lowest values in the realised network. In the
realised network, the nodes with a high Betweenness are distributed both
in the periphery and in the geographic centre of the network, indicating
a higher cohesion between distant nodes in the network promoted by these
"bridge" nodes. The connections between the pair of nodes , , ,{~ }and{~
}exhibited the highest edge Betweenness within the network. The
strongest connections, excluding self-loops, were , , ,{~ }and . It is
important to note that the high edge Betweenness of the 9-10 node pair
in the potential network was not observed in the realised network, while
node 10 formed a new bridge with node 7 (Fig. \ref{fig:fig6}).{~}

The four measures that highlight the importance of the nodes according
to their relationship to other nodes in a network (i.e., Eigenvector,
Hub and Authority score, and Page Rank) selected some of the same nodes
as critical in both the potential and realised networks. Node 1 was the
most important node in the realised network regarding Eigenvector,
Hub-score and Page Rank, and it ranked high in Authority-score, after
nodes 9 and 8 (Table S3, Fig. \ref{fig:fig4} and Fig. \ref{fig:fig7}). There is an almost total
coincidence in the classification of the nodes according to Page Rank
for the two networks (potential and realised), indicating that this
measure of centrality is not sensitive to the inclusion of demographic
information, or that it is a measure relative to egg production (similar
to RLR index).

\subsection{  Larval connectivity network in the simulated scenario}

In order to test the influence of node protection on the network
connectivity, we simulated a scenario where open-access nodes 3 and 7
are converted to no-take areas through the assignment of the biological
parameters observed in Cabrera MPA nodes, 10-12. Otherwise, no-take
nodes 1 and 9 are ``open to fisheries'' by the assignment of the average
biological parameters observed in open access areas. We explored the
structure of the network in this scenario and generally observed higher
scores of LR, RLR and SR indices (Fig. \ref{fig:fig3}) and of Strength, Out-strength
and In-strength for nodes 3 and 7, and higher connectivity measures for
node 7 (particularly Closeness and Betweenness; Figs. \ref{fig:fig4} and \ref{fig:fig6}, Table
S4). Otherwise, nodes 1 and 9 obtained similar scores. Overall, this
scenario obtained an improved connectivity throughout the nodes of the
realised larval connectivity network (e.g., see Authority score in
Tables S3 and S4 and Fig. \ref{fig:fig7}).

\subsection{Community detection}

\begin{figure}
	\centering
	\includegraphics[width=1\linewidth]{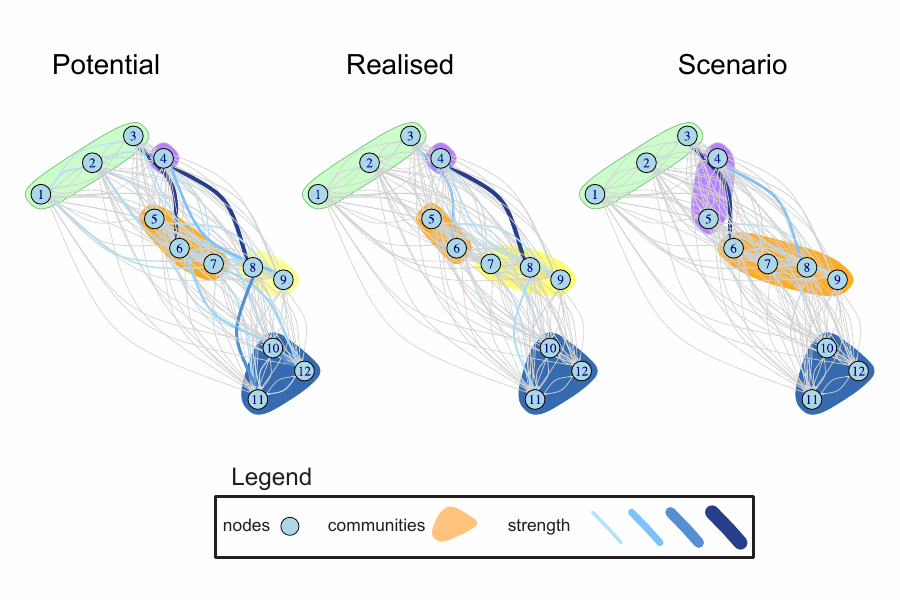}
	\caption{Graph-based community detection results for the potential, realised and
		scenario larval connectivity networks. The network communities were
		calculated finding the optimal{ }community structure by maximising the
		modularity measure over all possible partitions. Note that community
		here does not refer to biological communities but to a subset of nodes
		within the graph such that connections between the nodes are denser than
		connections with the rest of the network.}
	\label{fig:fig8}
\end{figure}

A total of 5 communities were detected in both the potential and
realised larval connectivity networks, whereas only four communities
were detected in the scenario network (Fig. \ref{fig:fig8}). The community formed by
nodes 1, 2 and 3 is persistent in all three networks, as is the
community formed by nodes 10, 11 and 12. However, nodes 4 to 9 are
shifting their community membership across the networks. Node 4, which
remains isolated in the potential and realised networks, is forming a
new community with node 5 in the simulated scenario.

\section{Discussion}

Larval transport is a key factor to consider in MPA network design.
Optimal sites regarding larval connectivity metrics are not always easy
to identify, particularly in complex scenarios constituted by intricate
geographical settings, heterogeneous habitats, or where strong
variations in physical ocean conditions occur. Our results show that, in
these settings, the combination of larval connectivity patterns in
complex marine systems and the network emergent properties (centrality
measures) from graph theory may optimise decisions on design and
management of MPA networks.

Graph theory allows spatially explicit representation of a complex
inter-connected ecological system {(Saunders et al. 2016)} because
different populations, represented as nodes, are connected by links that
represent population connectivity pathways {(Loro et al. 2015)}. As
shown in the present study, measures like Betweenness and Closeness are
good descriptors of MPA network properties, which is not explicit from
traditional cartography. The adequacy of these measurements depends on
the management or conservation targets and the resources to be managed.
For example, when managing a network of reserves, often the objective is
to maximise biological connectivity {(Andrello et al. 2015)}, either by
facilitating the mobility of adult organisms or by ensuring the
departure, arrival and mobility between nodes of larvae or propagules
(i.e., recruits). In this case, local flow patterns and the biological
features of the organisms, including their PLD and reproductive
behaviour, play a crucial role. In a broad sense, optimal sites for
conservation should meet at least three basic conditions: a) enhanced
larval production; b) increased larval recruitment; and c) high
connectivity amongst sites to ensure larval flow and avoid the isolation
of populations.

We demonstrated that applying graph theory enables to gain a deeper
insight into egg/larval dispersal network structure and function in a
complex system. For example, Betweenness centrality identified nodes
that have strategic importance for channelling large flow of propagules
(high Betweenness), as nodes from 5 to 8 that are in a hydrodynamically
active area close to a cape (Fig. 1). The incorporation of spatially
explicit information on reproductive output and graph theory analysis
enabled the identification of nodes acting as stepping-stones within the
complete network (Betweenness centrality), as well as the main paths
through which the larvae are transported (edge-Betweenness). For
example, a recent study has used Betweenness to identify key nodes that
act as gateways to dispersal in an MPA-OAA network in an archipelago in
the Western Indian Ocean{ (Gamoyo et al. 2019).} At the same time,
Closeness captures how close a node is to the ``central region'' of the
network. But here "central region" does not correspond to the
geographical centre but to the central source of larvae dissemination.
This "central region" can change from species to species, therefore, the
Closeness value for each node will change for each realised larval
connectivity network, when species-specific demographic information is
incorporated (Fig. 4). Hence, it is possible to identify the best and
worst connected nodes (using Closeness) independently of their spatial
location.

If egg production is uniform throughout the network, differences in
centrality measures should not be expected. However, demographic
differences between species in an ecosystem should be expected. For
example, in nature, individuals tend to aggregate where conditions are
more favourable. Marked differences in population reproductive
parameters (e.g., batch fecundity, maternal effects, etc.) have been
observed in \emph{S. scriba} at relatively short distances {(few
kilometres, Alós et al. 2013).} This spatial variability highlights the
necessity of including demographic parameters when estimating
connectivity. Other factors like larval behaviour are also relevant, yet
not always available. For example, vertical migration {(Ospina-Alvarez,
Parada, et al. 2012, Daigle et al. 2016)}; directionality in swimming in
response to the sound of waves breaking on shore {(Tolimieri et al.
2000, Montgomery et al. 2006) }or to chemical stimuli {(Sweatman 1988,
Paris, Atema, et al. 2013)}; and aggregation under floating objects
{(Shanks 1983, Tully \& Ceidigh 1989, Ohta \& Tachihara 2004)} have been
shown to modify the trajectories and final destination of fish larvae.
Our study relies on \emph{S. scriba}, due to availability of biological
data on the species in the study area, but the spatial optimization of
MPAs in other case studies should incorporate a set of species
representative of the species pool per habitat type in the region
{(Blanco et al. 2019)}.

Despite these limitations, the potential connectivity network is highly
relevant, as it can be used in absence of detailed biological
information of the key species, or species' pool, in the area of
interest. However, incorporating biological parameters of a key species
allowed us to obtain a more realistic representation of the connectivity
patterns and understanding of stock dynamics {(Morgan \& Fisher 2010,
Kell et al. 2015)}.The comparison of the two networks (potential vs.
realised) evidences changes in the relative importance of nodes as
stepping-stones. For example, in our realised connectivity network, node
5 emerged as a key node, while the importance of node 6, a key component
of the potential network, decreased drastically. In the case study this
is relevant, because node 5 is in fact a marine reserve (Fig. 1) that,
according to the centrality measures chosen, acts well as a bridge
between communities (Fig. 8) and as a node with high egg exportation
with subsequent high larval recruitment rates. We observed how
Betweenness and Closeness centrality ranked differently the important
nodes in the realised connectivity matrix. We also observed how the
bridge links between stepping-stone nodes and the classification of the
best-connected sites (according to their Closeness, Fig. 4) changed
between these networks, despite the case study was a small network
dominated by high retention and self-recruitment.{~}

MPA network design involves making decisions about the allocation of
protected areas, interspaced by areas open to fisheries (or under more
permissive fisheries restriction regimes). The alternative scenario
showed potential benefits for the network performance, particularly
improving the Betweenness in Palma Bay (nodes 2 to 5). Also, it achieved
a better balance of Out-strength and In-strength measures in the
southern cape area (node 7). Still, as revealed by Strength,
Out-strength and Betweenness, nodes 3 to 5 are crucial for the
performance of the network. However, node 7 gained relevance in
maintaining the connectivity with the more peripheral nodes, including
the Cabrera Archipelago MPA (nodes 10 to 12). By allocating MPAs in the
best performing sites in terms of larval connectivity, ecological
connectivity through the network is improved and, therefore, the
resilience of the whole spawner-recruit system is strengthened. We also
show how closing sites to fishing activities leads to different patterns
of connectivity or groupings of sites. Management decisions based on
centrality measures can increase network resilience by allowing, through
simple analysis, decision-making in networks composed of several
management units (Fig. 8).

The graph analysis performed over the realised connectivity network
eases the understanding of some ecological processes {(e.g., gene flow,
colonization, invasion; Moilanen 2011)} and demonstrates that the
incorporation of biological data provides an additional and valuable
source of information to assist management decisions. A key advantage of
graph analysis is the information provided by centrality measures, which
allow to identify the most critical nodes acting as bridges between
communities. We encourage scientists and decision-makers to choose
within the set of centrality measures derived from graph theory, those
that best define their case study according to their ecological
significance. This will assist the adoption of the concept
``reproductive system'' by {Lowerre-Barbieri et al.} {(2016)} as a more
realistic framework to analyse population resilience through time and
produce a more detailed view of the connectivity patterns in
meta-populations of marine species.

\section*{Acknowledgments}

This study was supported by the research project Phenofish (grant number
CTM2015-69126-C2-1-R) funded by the Spanish Ministry of Economy and
Competiveness. AO was supported by H2020 Marie Skłodowska-Curie Actions
MSCA-IF-2016 (Project ID: 746361). SJ was supported by H2020- Marie
Skłodowska-Curie Actions MSCA-IF-2016 (Project ID: 743545). JA was
supported by a Ramón y Cajal Grant funded by the Spanish Ministry of
Science, Innovation and Universities (Grant No. RYC2018-024488-I). GFB
was funded by a fellowship (FPI-INIA) (CPD2015) from the National
Institute for Agricultural and Food Research and Technology (INIA).

\section*{References}
\begin{scriptsize}
\begin{list}{}{}
	\item Alonso-Fernández A, Alós J, Grau A, Domínguez-Petit R, Saborido-Rey F
(2011) The Use of Histological Techniques to Study the Reproductive
Biology of the Hermaphroditic Mediterranean Fishes Coris julis, Serranus
scriba, and Diplodus annularis. Marine and Coastal Fisheries 3:145--159

	\item Alós J, Alonso-Fernández A, Catalán IA, Palmer M, Lowerre-Barbieri S
(2013) Reproductive output traits of the simultaneous hermaphrodite
\emph{Serranus scriba} in the western Mediterranean. Sci Mar 77:331--340

	\item Alós J, Palmer M, Catalán IA, Alonso-Fernández A, Basterretxea G, Jordi
A, Buttay L, Morales-Nin B, Arlinghaus R (2014) Selective exploitation
of spatially structured coastal fish populations by recreational anglers
may lead to evolutionary downsizing of adults. Mar Ecol Prog Ser
503:219--233

	\item Andrello M, Jacobi MN, Manel S, Thuiller W, Mouillot D (2015) Extending
networks of protected areas to optimize connectivity and population
growth rate. Ecography 38:273--282

	\item Armsworth PR (2002) Recruitment limitation, population regulation, and
larval connectivity in reef fish metapopulations. Ecology 83:1092--1104

	\item Basterretxea G, Jordi A, Catalán IA, Sabatés A (2012) Model-based
assessment of local-scale fish larval connectivity in a network of
marine protected areas. Fish Oceanogr 21:291--306

	\item Blanco M, Ospina-Alvarez A, Navarrete SA, Fernández M (2019) Influence
of larval traits on dispersal and connectivity patterns of two exploited
marine invertebrates in central Chile. Mar Ecol Prog Ser 612:43--64

	\item Botsford LW, Brumbaugh DR, Grimes C, Kellner JB, Largier J, O'Farrell
MR, Ralston S, Soulanille E, Wespestad V (2008) Connectivity,
sustainability, and yield: bridging the gap between conventional
fisheries management and marine protected areas. Rev Fish Biol Fish
19:69--95

	\item Botsford LW, Micheli F, Hastings A (2003) Principles for the design of
marine reserves. Ecol Appl 13:25--31

	\item Botsford LW, White JW, Coffroth MA, Paris CB, Planes S, Shearer TL,
Thorrold SR, Jones GP (2009) Connectivity and resilience of coral reef
metapopulations in marine protected areas: matching empirical efforts to
predictive needs. Coral Reefs 28:327--337

	\item Brandes U, Delling D, Gaertler M, Görke R, Hoefer M, Nikoloski Z, Wagner
D (2008) On modularity clustering. TKDE 20:172--188

	\item Catalán IA, Macías D, Solé J, Ospina-Alvarez A, Ruíz J (2013) Stay off
the motorway: resolving the pre-recruitment life history dynamics of the
European anchovy in the SW Mediterranean through a spatially-explicit
individual-based model (SEIBM). Prog Oceanogr 111:140--153

	\item Chang W (2012) R Graphics Cookbook, 1st edn. O'Reilly, Sebastopol

	\item Conklin EE, Neuheimer AB, Toonen RJ (2018) Modeled larval connectivity
of a multi-species reef fish and invertebrate assemblage off the coast
of Moloka`i, Hawai`i. PeerJ 6:e5688--32

	\item Cowen RK, Sponaugle S (2009) Larval dispersal and marine population
connectivity. Annu Rev Mar Sci 1:443--466

	\item Cowen RK, Paris CB, Srinivasan A (2006) Scaling of connectivity in
marine populations. Science 311:522--527

	\item Csardi G, Nepusz T (2006) The igraph software package for complex
network research. InterJournal Complex Systems:1695

	\item Daigle RM, Chassé J, Metaxas A (2016) The relative effect of behaviour
in larval dispersal in a low energy embayment. Prog Oceanogr 144:93--117

	\item Dale MRT, Fortin MJ (2010) From Graphs to Spatial Graphs. Annu Rev Ecol
Evol S 41:21--38

	\item Donahue MJ, Karnauskas M, Toews C, Paris CB (2015) Location isn't
everything: Timing of spawning aggregations optimizes larval
replenishment. PLoS ONE 10:e0130694

	\item Fogarty MJ, Botsford L (2007) Population connectivity and spatial
management of marine fisheries. Oceanography 20:112--123

	\item Freeman LC (1978) Centrality in social networks conceptual
clarification. Social Networks 1:215--239

	\item Friesen SK, Martone R, Rubidge E, Baggio JA, Ban NC (2019) An approach
to incorporating inferred connectivity of adult movement into marine
protected area design with limited data. Ecol Appl 29:e01890--14

	\item Gaines SD, White C, Carr MH, Palumbi SR (2010) Designing marine reserve
networks for both conservation and fisheries management. P Natl Acad Sci
USA 107:18286--18293

	\item Galarza JA, Carreras-Carbonell J, Macpherson E, Pascual M, Roques S,
Turner GF, Rico C (2009) The influence of oceanographic fronts and
early-life-history traits on connectivity among littoral fish species. P
Natl Acad Sci USA 106:1473--1478

	\item Gamoyo M, Obura D, Reason CJC (2019) Estimating Connectivity Through
Larval Dispersal in the Western Indian Ocean. J Geophys Res Biogeosci
124:2446--2459

	\item Girvan M, Newman M (2002) Community structure in social and biological
networks. P Natl Acad Sci USA 99:7821--7826

	\item Guimera R, Nunes Amaral LA (2005) Functional cartography of complex
metabolic networks. Nature 433:895--900

	\item Gwinn DC, Allen MS, Johnston FD, Brown P, Todd CR, Arlinghaus R (2015)
Rethinking length‐based fisheries regulations: the value of protecting
old and large fish with harvest slots. Fish and Fisheries 16:259--281

	\item Hedgecock D, Barber PH, Edmands S (2007) Genetic approaches to measuring
connectivity. Oceanography 20:70--79

	\item Henry L-A, Mayorga-Adame CG, Fox AD, Polton JA, Ferris JS, McLellan F,
McCabe C, Kutti T, Roberts JM (2018) Ocean sprawl facilitates dispersal
and connectivity of protected species. Sci Rep 8:1--11

	\item Hinckley S, Hermann A, Mier K, Megrey BA (2001) Importance of spawning
location and timing to successful transport to nursery areas: a
simulation study of Gulf of Alaska walleye pollock. ICES J Mar Sci
58:1042--1052

	\item Hixon MA, Johnson DW, Sogard SM (2014) BOFFFFs: on the importance of
conserving old-growth age structure in fishery populations. ICES J Mar
Sci 71:2171--2185

	\item Jacobi MN, Jonsson PR (2011) Optimal networks of nature reserves can be
found through eigenvalue perturbation theory of the connectivity matrix.
Ecol Appl 21:1861--1870

	\item Kaplan DM, Cuif M, Fauvelot C, Vigliola L, Nguyen-Huu T, Tiavouane J,
Lett C (2017) Uncertainty in empirical estimates of marine larval
connectivity. ICES J Mar Sci 74:1723--1734

	\item Kell LT, Nash RDM, Dickey-Collas M, Mosqueira I, Szuwalski C (2015) Is
spawning stock biomass a robust proxy for reproductive potential? Fish
and Fisheries:doi: 10.1111--faf.12131

	\item Kininmonth S, Weeks R, Abesamis RA, Bernardo LPC, Beger M, Treml EA,
Williamson D, Pressey RL (2018) Strategies in scheduling marine
protected area establishment in a network system. Ecol Appl 1695:1--10

	\item Kininmonth SJ, De'ath G, Possingham HP (2009) Graph theoretic topology
of the Great but small Barrier Reef world. Theor Ecol 3:75--88

	\item Kinlan BP, Gaines SD (2003) Propagule dispersal in marine and
terrestrial environments: a community perspective. Ecology 84:2007--2020

	\item Kough AS, Paris CB (2015) The influence of spawning periodicity on
population connectivity. Coral Reefs 34:753--757

	\item Lagabrielle E, Crochelet E, Andrello M, Schill SR, Arnaud-Haond S,
Alloncle N, Ponge B (2014) Connecting MPAs -- eight challenges for
science and management. Aquat Conserv 24:94--110

	\item Lett C, Nguyen-Huu T, Cuif M, Saenz-Agudelo P, Kaplan DM (2015) Linking
local retention, self-recruitment, and persistence in marine
metapopulations. Ecology 96:2236--2244

	\item Loro M, Ortega E, Arce RM, Geneletti D (2015) Ecological connectivity
analysis to reduce the barrier effect of roads. An innovative
graph-theory approach to define wildlife corridors with multiple paths
and without bottlenecks. Landscape and Urban Planning 139:149--162

	\item Lowerre-Barbieri S, DeCelles G, Pepin P, Catalán IA, Muhling B, Erisman
B, Cadrin SX, Alós J, Ospina-Alvarez A, Stachura MM, Tringali MD,
Burnsed SW, Paris CB (2016) Reproductive resilience: a paradigm shift in
understanding spawner‐recruit systems in exploited marine fish. Fish and
Fisheries 18:285--312

	\item Lubchenco J, Palumbi SR, Gaines SD, Andelman S (2003) Plugging a Hole in
the Ocean: The Emerging Science of Marine Reserves. Ecol Appl 13:S3--S7

	\item Macpherson E, Raventós N (2006) Relationship between pelagic larval
duration and geographic distribution in Mediterranean littoral fishes.
Mar Ecol Prog Ser 327:257--265

	\item March D, Palmer M, Alós J, Grau A, Cardona F (2010) Short-term
residence, home range size and diel patterns of the painted comber
Serranus scriba in a temperate marine reserve. Mar Ecol Prog Ser
400:195--206

	\item Moilanen A (2011) On the limitations of graph‐theoretic connectivity in
spatial ecology and conservation. J Appl Ecol 48:1543--1547

	\item Montgomery JC, Jeffs A, Simpson SD, Meekan M, Tindle C (2006) Sound as
an orientation cue for the pelagic larvae of reef fishes and decapod
crustaceans. Adv Mar Biol 51:143--196

	\item Morgan SG (2014) Behaviorally mediated larval transport in upwelling
systems. Advances in Oceanography 2014:1--17

	\item Morgan SG, Fisher JL (2010) Larval behavior regulates nearshore
retention and offshore migration in an upwelling shadow and along the
open coast. Mar Ecol Prog Ser 404:109--126

	\item Norcross BL, Shaw RF (1984) Oceanic and estuarine transport of fish eggs
and larvae: A review. Trans Am Fish Soc 113:153--165

	\item Ohta I, Tachihara K (2004) Larval development and food habits of the
marbled parrotfish, Leptoscarus vaigiensis , associated with drifting
algae. Ichthyol Res 51:63--69

	\item Ospina-Alvarez A, Catalán IA, Bernal M, Roos D, Palomera I (2015) From
egg production to recruits: Connectivity and inter-annual variability in
the recruitment patterns of European anchovy in the northwestern
Mediterranean. Prog Oceanogr 138:431--447

	\item Ospina-Alvarez A, de Juan S, Davis KJ, González C, Fernández M,
Navarrete SA (2020) Integration of biophysical connectivity in the
spatial optimization of coastal ecosystem services. Sci Total Environ
733:139367

	\item Ospina-Alvarez A, Palomera I, Parada C (2012) Changes in egg buoyancy
during development and its effects on the vertical distribution of
anchovy eggs. Fish Res 117:86--95

	\item Ospina-Alvarez A, Parada C, Palomera I (2012) Vertical migration effects
on the dispersion and recruitment of European anchovy larvae: From
spawning to nursery areas. Ecol Model 231:65--79

	\item Ospina-Alvarez A, Weidberg N, Aiken CM, Navarrete SA (2018) Larval
transport in the upwelling ecosystem of central Chile: The effects of
vertical migration, developmental time and coastal topography on
recruitment. Prog Oceanogr 168:82--99

	\item Paris CB, Atema J, Irisson J-O, Kingsford M, Gerlach G, Guigand CM
(2013) Reef odor: A wake up call for navigation in reef fish larvae (C
Fulton, Ed.). PLoS ONE 8:e72808

	\item Paris CB, Helgers J, van Sebille E, Srinivasan A (2013) Connectivity
Modeling System: A probabilistic modeling tool for the multi-scale
tracking of biotic and abiotic variability in the ocean. Environ Modell
Softw 42:47--54

	\item Pineda J (2000) Linking larval settlement to larval transport:
assumptions, potentials, and pitfalls. Oceanography of the Eastern
Pacific 1:84--105

	\item R Core Team (2019) R: A Language and Environment for Statistical
Computing. Vienna, Austria

	\item Reichardt J, Bornholdt S (2006) Statistical Mechanics of Community
Detection. arXiv cond-mat.dis-nn

	\item Saunders MI, Brown CJ, Foley MM, Febria CM, Albright R, Mehling MG,
Kavanaugh MT, Burfeind DD (2016) Human impacts on connectivity in marine
and freshwater ecosystems assessed using graph theory: a review. Mar
Freshwater Res 67:277--290

	\item Shanks AL (1983) Surface slicks associated with tidally forced internal
waves may transport pelagic larvae of benthic invertebrates and fishes
shoreward. Mar Ecol Prog Ser 13:311--315

	\item Siegel DA, Kinlan BP, Gaylord B, Gaines SD (2003) Lagrangian
descriptions of marine larval dispersion. Mar Ecol Prog Ser 260:83--96

	\item Siegel DA, Mitarai S, Costello CJ, Gaines SD, Kendall BE, Warner RR,
Winters KB (2008) The stochastic nature of larval connectivity among
nearshore marine populations. P Natl Acad Sci USA 105:8974--8979

	\item Strathmann RR, Hughes TR, Kuris AM, Lindeman KC, Morgan SG, Pandolfi JM,
Warner RR (2002) Evolution of self-recruitment and its consequences for
marine populations. Bulletin of Marine Science -Miami- 70

	\item Sweatman H (1988) Field evidence that settling coral reef fish larvae
detect resident fishes using dissolved chemical cues. J Exp Mar Biol
Ecol 124:163--174

	\item Thorrold SR, Zacherl DC, Levin LA (2007) Population connectivity and
larval dispersal using geochemical signatures in calcified structures.
Oceanography 20:80--89

	\item Tolimieri N, Jeffs A, Montgomery JC (2000) Ambient sound as a cue for
navigation by the pelagic larvae of reef fishes. Mar Ecol Prog Ser
207:219--224

	\item Travis JMJ, Dytham C (1998) The evolution of dispersal in a
metapopulation: a spatially explicit, individual-based model. Proc R Soc
Lond, B, Biol Sci 265:17--23

	\item Treml EA, Halpin PN, Urban DL, Pratson LF (2008) Modeling population
connectivity by ocean currents, a graph-theoretic approach for marine
conservation. Landscape Ecol 23:19--36

	\item Tully O, Ceidigh PÓ (1989) The ichthyoneuston of Galway Bay (west of
Ireland). II. Food of post-larval and juvenile neustonic and
pseudo-neustonic fish. Mar Ecol Prog Ser 51:301--310

	\item Walford LA (1938) Effect of currents on distribution and survival of the
eggs and larvae of the haddock (\emph{Melanogrammius aeglefinus}) on
Georges bank. Bulletin of the Bureau of Fisheries 49:1--73

	\item Watson JR, Kendall BE, Siegel DA, Mitarai S (2012) Changing seascapes,
stochastic connectivity, and marine metapopulation dynamics. Am Nat
180:99--112

	\item Watson JR, Mitarai S, Siegel DA, Caselle JE, Dong C, McWilliams JC
(2010) Realized and potential larval connectivity in the Southern
California Bight. Mar Ecol Prog Ser 401:31--48

	\item Werner FE, Cowen RK, Paris CB (2007) Coupled biological and physical
models. Oceanography

	\item Wickham H (2016) ggplot2: Elegant Graphics for Data Analysis. Springer
International Publishing, New York
\end{list}
\end{scriptsize}

\appendix
\label{sec:Appendix}



\section*{Supplementary material}

# Supplementary Material

*Ms. entitled "MPA network design based on graph theory and emergent properties of larval dispersal"*

*By: Andres Ospina-Alvarez, Silvia de Juan, Josep Alós, Gotzon Basterretxea, Alexandre Alonso-Fernández, Guillermo Follana-Berná, Miquel Palmer, and Ignacio A. Catalán.*

Table S1: Pelagic larval duration (mean and SD) in days and range of breeding season of some littoral species in the Mediterranean Sea and inhabiting the Balearic Islands. Adapted from Macpherson and Raventós (2006)[1]. The last row includes the PLD and range of months used in the Lagrangian transport simulations by Basterretxea et al. (2012)[2].

| Common name | Scientific name | mean PLD | SD | Breeding season |
|---|---|---|---|---|
| Cardinal fish | *Apogon imberbis* | 21.3 | 1.4 | July to September |
| Adriatic blenny | *Lipophrys adriaticus* | 23.0 | 1.7 | April to August |
| Mystery blenny | *Parablennius incognitus* | 23.8 | 1.6 | February to July |
| Bucchich's goby | *Gobius bucchichi* | 19.2 | 1.0 | April to June |
| Goldsinny wrasse | *Ctenolabrus rupestris* | 20.9 | 2.0 | January to July |
| Brown meagre | *Sciaena umbra* | 22.5 | 0.7 | March to August |
| Dusky groupe | *Epinephelus marginatus* | 24.6 | 1.3 | July to September |
| Comber | *Serranus cabrilla* | 24.3 | 1.8 | April to July |
| Brown comber | *Serranus hepatus* | 18.0 | 0.9 | March to August |
| Annular seabream | *Diplodus annularis* | 18.3 | 1.4 | May to June |
| Zebra seabream | *Diplodus cervinus* | 18.3 | 1.5 | January to April |
| Straightnose pipefish | *Nerophis ophidion* | 21.5 | 0.7 | May to August |
| Tub gurnard | *Chelidonichthys lucerna* | 19.1 | 1.2 | October to April |
| Red-black triplefin | *Tripterygion tripteronotum* | 18.4 | 2.8 | April to August |
| Lagrangian transport simu. | NA | 21.0 | NA | March to August |

Table S2. Index or centrality measure, normalized* value and sorted nodes for the potential larval connectivity network. Local Retention (LR), Relative Local Retention (RLR), Self-Recruitment (SR), Strength (STR), Out-Strength (OST), In-Strength (IST), Closeness (CLO), Betweenness (BET), Eigenvector centrality (EIV), Hub score (HUB), Authority score (AUT) and Page Rank (PAR).
* The normalization was performed using the following formula: (x - min) / (max - min).

| Category | LR | | | RLR | | | SR | | | STR | | | OST | | | IST | | |
|---|---|---|---|---|---|---|---|---|---|---|---|---|---|---|---|---|---|---|
| | Index value | Norm. value | Sorted nodes | Index value | Norm. value | Sorted nodes | Index value | Norm. value | Sorted nodes | Cent. Mea. | Norm. value | Sorted nodes | Cent. Mea. | Norm. value | Sorted nodes | Cent. Mea. | Norm. value | Sorted nodes |
| Highest | 0.082 | 1.000 | 3 | 0.820 | 1.000 | 3 | 0.652 | 1.000 | 11 | 0.231 | 1.000 | 4 | 0.100 | 1.000 | 3 | 0.133 | 1.000 | 4 |
| | 0.077 | 0.918 | 4 | 0.787 | 0.942 | 4 | 0.632 | 0.930 | 3 | 0.231 | 1.000 | 3 | 0.098 | 0.949 | 4 | 0.131 | 0.975 | 3 |
| High | 0.063 | 0.700 | 11 | 0.714 | 0.814 | 11 | 0.627 | 0.915 | 12 | 0.192 | 0.718 | 10 | 0.091 | 0.788 | 8 | 0.104 | 0.700 | 10 |
| | 0.059 | 0.625 | 10 | 0.672 | 0.740 | 10 | 0.613 | 0.866 | 1 | 0.186 | 0.678 | 11 | 0.089 | 0.751 | 11 | 0.097 | 0.626 | 11 |
| | 0.054 | 0.549 | 12 | 0.640 | 0.686 | 12 | 0.588 | 0.778 | 9 | 0.177 | 0.614 | 8 | 0.087 | 0.719 | 10 | 0.091 | 0.554 | 6 |
| Medium | 0.047 | 0.434 | 8 | 0.514 | 0.465 | 8 | 0.581 | 0.753 | 4 | 0.176 | 0.608 | 6 | 0.086 | 0.692 | 5 | 0.087 | 0.513 | 8 |
| | 0.036 | 0.272 | 9 | 0.508 | 0.455 | 9 | 0.562 | 0.685 | 10 | 0.171 | 0.568 | 7 | 0.086 | 0.691 | 7 | 0.086 | 0.504 | 12 |
| | 0.033 | 0.223 | 6 | 0.432 | 0.321 | 1 | 0.538 | 0.601 | 8 | 0.170 | 0.561 | 12 | 0.086 | 0.685 | 6 | 0.085 | 0.493 | 7 |
| Low | 0.032 | 0.207 | 7 | 0.387 | 0.242 | 6 | 0.473 | 0.373 | 2 | 0.134 | 0.304 | 5 | 0.084 | 0.648 | 12 | 0.062 | 0.251 | 9 |
| | 0.023 | 0.066 | 1 | 0.373 | 0.219 | 7 | 0.446 | 0.280 | 5 | 0.133 | 0.297 | 9 | 0.071 | 0.377 | 9 | 0.048 | 0.105 | 5 |
| | 0.021 | 0.037 | 5 | 0.280 | 0.056 | 2 | 0.379 | 0.046 | 7 | 0.108 | 0.119 | 2 | 0.068 | 0.303 | 2 | 0.040 | 0.025 | 2 |
| Lowest | 0.019 | 0.000 | 2 | 0.248 | 0.000 | 5 | 0.366 | 0.000 | 6 | 0.091 | 0.000 | 1 | 0.054 | 0.000 | 1 | 0.038 | 0.000 | 1 |

| Category | CLO | | | BET | | | EIV | | | HUB | | | AUT | | | PAR | | |
|---|---|---|---|---|---|---|---|---|---|---|---|---|---|---|---|---|---|---|
| | Cent. Mea. | Norm. value | Sorted nodes | Cent. Mea. | Norm. value | Sorted nodes | Cent. Mea. | Norm. value | Sorted nodes | Cent. Mea. | Norm. value | Sorted nodes | Cent. Mea. | Norm. value | Sorted nodes | Cent. Mea. | Norm. value | Sorted nodes |
| Highest | 0.044 | 1.000 | 6 | 57.000 | 1.000 | 6 | 1.000 | 1.000 | 3 | 1.000 | 1.000 | 11 | 1.000 | 1.000 | 1 | 0.145 | 1.000 | 1 |
| | 0.043 | 0.985 | 7 | 55.000 | 0.965 | 7 | 0.878 | 0.850 | 9 | 0.972 | 0.968 | 10 | 0.770 | 0.729 | 2 | 0.124 | 0.806 | 9 |
| High | 0.040 | 0.875 | 5 | 35.000 | 0.614 | 8 | 0.742 | 0.683 | 4 | 0.912 | 0.902 | 9 | 0.763 | 0.722 | 3 | 0.113 | 0.701 | 2 |
| | 0.031 | 0.536 | 2 | 28.000 | 0.491 | 5 | 0.735 | 0.675 | 1 | 0.904 | 0.894 | 12 | 0.444 | 0.346 | 5 | 0.099 | 0.579 | 3 |
| | 0.030 | 0.508 | 8 | 27.000 | 0.474 | 10 | 0.727 | 0.666 | 12 | 0.569 | 0.521 | 8 | 0.419 | 0.316 | 4 | 0.093 | 0.522 | 8 |
| Medium | 0.029 | 0.474 | 1 | 24.000 | 0.421 | 9 | 0.701 | 0.634 | 11 | 0.498 | 0.442 | 3 | 0.362 | 0.250 | 9 | 0.085 | 0.450 | 12 |
| | 0.028 | 0.423 | 9 | 16.000 | 0.281 | 4 | 0.656 | 0.579 | 10 | 0.471 | 0.412 | 4 | 0.291 | 0.166 | 8 | 0.068 | 0.287 | 5 |
| | 0.026 | 0.341 | 4 | 12.000 | 0.211 | 2 | 0.605 | 0.516 | 2 | 0.207 | 0.119 | 7 | 0.223 | 0.087 | 12 | 0.064 | 0.252 | 11 |
| Low | 0.023 | 0.235 | 3 | 2.000 | 0.035 | 3 | 0.601 | 0.511 | 8 | 0.125 | 0.027 | 2 | 0.163 | 0.016 | 7 | 0.062 | 0.232 | 10 |
| | 0.018 | 0.062 | 10 | 1.000 | 0.018 | 11 | 0.353 | 0.207 | 5 | 0.120 | 0.022 | 6 | 0.162 | 0.015 | 11 | 0.062 | 0.231 | 4 |
| | 0.018 | 0.041 | 12 | 1.000 | 0.018 | 1 | 0.268 | 0.104 | 7 | 0.108 | 0.009 | 1 | 0.156 | 0.008 | 10 | 0.050 | 0.126 | 7 |
| Lowest | 0.017 | 0.000 | 11 | 0.000 | 0.000 | 12 | 0.184 | 0.000 | 6 | 0.100 | 0.000 | 5 | 0.150 | 0.000 | 6 | 0.036 | 0.000 | 6 |

*Table S3. Index or centrality measure, normalized\* value and sorted nodes for the realised larval connectivity network. Local Retention (LR), Relative Local Retention (RLR), Self-Recruitment (SR), Strength (STR), Out-Strength (OST), In-Strength (IST), Closeness (CLO), Betweenness (BET), Eigenvector centrality (EIV), Hub score (HUB), Authority score (AUT) and Page Rank (PAR).*
*\* The normalization was performed using the following formula: (x - min) / (max - min).*

| Category | LR Index value | LR Norm. value | LR Sorted nodes | RLR Index value | RLR Norm. value | RLR Sorted nodes | SR Index value | SR Norm. value | SR Sorted nodes | STR Cent. Mea. | STR Norm. value | STR Sorted nodes | OST Cent. Mea. | OST Norm. value | OST Sorted nodes | IST Cent. Mea. | IST Norm. value | IST Sorted nodes |
|---|---|---|---|---|---|---|---|---|---|---|---|---|---|---|---|---|---|---|
| Highest | 0.191 | 1.000 | 4 | 0.825 | 1.000 | 3 | 0.7513 | 1.000 | 4 | 0.497 | 1.000 | 4 | 0.244 | 1.000 | 4 | 0.254 | 1.000 | 4 |
|  | 0.100 | 0.521 | 3 | 0.782 | 0.926 | 4 | 0.6905 | 0.911 | 12 | 0.286 | 0.555 | 3 | 0.154 | 0.626 | 5 | 0.164 | 0.621 | 3 |
| High | 0.063 | 0.323 | 10 | 0.713 | 0.806 | 11 | 0.6645 | 0.873 | 5 | 0.211 | 0.399 | 5 | 0.121 | 0.491 | 3 | 0.095 | 0.325 | 10 |
|  | 0.051 | 0.261 | 8 | 0.668 | 0.729 | 10 | 0.6622 | 0.870 | 10 | 0.188 | 0.351 | 10 | 0.100 | 0.404 | 8 | 0.078 | 0.256 | 8 |
|  | 0.051 | 0.259 | 12 | 0.634 | 0.670 | 12 | 0.649 | 0.850 | 8 | 0.179 | 0.331 | 8 | 0.094 | 0.376 | 10 | 0.073 | 0.234 | 12 |
| Medium | 0.038 | 0.194 | 5 | 0.506 | 0.448 | 8 | 0.6096 | 0.793 | 3 | 0.153 | 0.276 | 12 | 0.080 | 0.317 | 12 | 0.067 | 0.209 | 6 |
|  | 0.026 | 0.128 | 9 | 0.494 | 0.427 | 9 | 0.536 | 0.685 | 9 | 0.109 | 0.185 | 7 | 0.065 | 0.256 | 2 | 0.061 | 0.182 | 7 |
|  | 0.019 | 0.091 | 11 | 0.448 | 0.347 | 1 | 0.5261 | 0.670 | 2 | 0.101 | 0.166 | 2 | 0.052 | 0.202 | 9 | 0.057 | 0.168 | 5 |
| Low | 0.019 | 0.091 | 2 | 0.393 | 0.250 | 6 | 0.3843 | 0.463 | 11 | 0.100 | 0.165 | 9 | 0.049 | 0.188 | 7 | 0.049 | 0.131 | 11 |
|  | 0.018 | 0.086 | 7 | 0.368 | 0.208 | 7 | 0.2942 | 0.331 | 7 | 0.079 | 0.121 | 6 | 0.026 | 0.095 | 11 | 0.048 | 0.127 | 9 |
|  | 0.005 | 0.016 | 6 | 0.287 | 0.068 | 2 | 0.0869 | 0.028 | 1 | 0.075 | 0.113 | 11 | 0.012 | 0.034 | 6 | 0.036 | 0.074 | 2 |
| Lowest | 0.002 | 0.000 | 1 | 0.248 | 0.000 | 5 | 0.068 | 0.000 | 6 | 0.022 | 0.000 | 1 | 0.003 | 0.000 | 1 | 0.018 | 0.000 | 1 |

| Category | CLO Cent. Mea. | CLO Norm. value | CLO Sorted nodes | BET Cent. Mea. | BET Norm. value | BET Sorted nodes | EIV Cent. Mea. | EIV Norm. value | EIV Sorted nodes | HUB Cent. Mea. | HUB Norm. value | HUB Sorted nodes | AUT Cent. Mea. | AUT Norm. value | AUT Sorted nodes | PAR Cent. Mea. | PAR Norm. value | PAR Sorted nodes |
|---|---|---|---|---|---|---|---|---|---|---|---|---|---|---|---|---|---|---|
| Highest | 0.055 | 1.000 | 5 | 57.000 | 1.000 | 5 | 1.000 | 1.000 | 1 | 1.000 | 1.000 | 1 | 1.000 | 1.000 | 9 | 0.139 | 1.000 | 1 |
|  | 0.040 | 0.680 | 4 | 44.000 | 0.772 | 7 | 0.747 | 0.661 | 11 | 0.809 | 0.805 | 11 | 0.773 | 0.684 | 8 | 0.130 | 0.905 | 9 |
| High | 0.033 | 0.550 | 3 | 32.000 | 0.561 | 4 | 0.665 | 0.551 | 9 | 0.257 | 0.240 | 3 | 0.687 | 0.564 | 1 | 0.109 | 0.707 | 2 |
|  | 0.030 | 0.472 | 8 | 30.000 | 0.526 | 10 | 0.503 | 0.334 | 3 | 0.251 | 0.234 | 6 | 0.657 | 0.522 | 3 | 0.102 | 0.635 | 3 |
|  | 0.028 | 0.436 | 7 | 20.000 | 0.351 | 2 | 0.431 | 0.237 | 8 | 0.248 | 0.231 | 9 | 0.618 | 0.468 | 5 | 0.092 | 0.545 | 8 |
| Medium | 0.027 | 0.411 | 2 | 18.000 | 0.316 | 3 | 0.405 | 0.203 | 12 | 0.229 | 0.211 | 12 | 0.569 | 0.400 | 2 | 0.085 | 0.479 | 12 |
|  | 0.025 | 0.370 | 9 | 14.000 | 0.246 | 8 | 0.351 | 0.130 | 2 | 0.188 | 0.169 | 10 | 0.495 | 0.297 | 12 | 0.069 | 0.323 | 5 |
|  | 0.018 | 0.244 | 10 | 0.000 | 0.000 | 9 | 0.325 | 0.096 | 6 | 0.120 | 0.100 | 4 | 0.435 | 0.213 | 4 | 0.064 | 0.272 | 11 |
| Low | 0.016 | 0.186 | 6 | 0.000 | 0.000 | 6 | 0.317 | 0.085 | 10 | 0.099 | 0.079 | 8 | 0.424 | 0.198 | 7 | 0.063 | 0.262 | 10 |
|  | 0.015 | 0.175 | 12 | 0.000 | 0.000 | 12 | 0.295 | 0.056 | 5 | 0.087 | 0.066 | 2 | 0.364 | 0.114 | 10 | 0.062 | 0.256 | 4 |
|  | 0.013 | 0.134 | 11 | 0.000 | 0.000 | 11 | 0.288 | 0.045 | 4 | 0.080 | 0.059 | 7 | 0.322 | 0.055 | 11 | 0.049 | 0.130 | 7 |
| Lowest | 0.007 | 0.000 | 1 | 0.000 | 0.000 | 1 | 0.254 | 0.000 | 7 | 0.022 | 0.000 | 5 | 0.282 | 0.000 | 6 | 0.036 | 0.000 | 6 |

*Table S4. Index or centrality measure, normalized\* value and sorted nodes for the larval connectivity network scenario. Local Retention (LR), Relative Local Retention (RLR), Self-Recruitment (SR), Strength (STR), Out-Strength (OST), In-Strength (IST), Closeness (CLO), Betweenness (BET), Eigenvector centrality (EIV), Hub score (HUB), Authority score (AUT) and Page Rank (PAR).*
*\* The normalization was performed using the following formula: (x - min) / (max - min).*

| Category | LR Index value | LR Norm. value | LR Sorted nodes | RLR Index value | RLR Norm. value | RLR Sorted nodes | SR Index value | SR Norm. value | SR Sorted nodes | STR Cent. Mea. | STR Norm. value | STR Sorted nodes | OST Cent. Mea. | OST Norm. value | OST Sorted nodes | IST Cent. Mea. | IST Norm. value | IST Sorted nodes |
|---|---|---|---|---|---|---|---|---|---|---|---|---|---|---|---|---|---|---|
| Highest | 0.257 | 1.000 | 3 | 0.825 | 1.000 | 3 | 0.8464 | 1.000 | 3 | 0.616 | 1.000 | 3 | 0.312 | 1.000 | 3 | 0.304 | 1.000 | 3 |
| | 0.136 | 0.527 | 4 | 0.782 | 0.926 | 4 | 0.6669 | 0.783 | 12 | 0.377 | 0.602 | 4 | 0.174 | 0.555 | 4 | 0.204 | 0.652 | 4 |
| High | 0.046 | 0.178 | 7 | 0.713 | 0.806 | 11 | 0.6669 | 0.783 | 4 | 0.200 | 0.307 | 7 | 0.125 | 0.399 | 7 | 0.075 | 0.207 | 7 |
| | 0.045 | 0.172 | 10 | 0.668 | 0.729 | 10 | 0.6327 | 0.741 | 10 | 0.158 | 0.237 | 5 | 0.110 | 0.350 | 5 | 0.071 | 0.193 | 6 |
| | 0.036 | 0.140 | 8 | 0.634 | 0.670 | 12 | 0.6159 | 0.721 | 7 | 0.137 | 0.203 | 10 | 0.072 | 0.227 | 8 | 0.071 | 0.193 | 10 |
| Medium | 0.036 | 0.139 | 12 | 0.506 | 0.448 | 8 | 0.5806 | 0.678 | 8 | 0.134 | 0.197 | 8 | 0.067 | 0.212 | 10 | 0.062 | 0.164 | 8 |
| | 0.027 | 0.105 | 5 | 0.494 | 0.427 | 9 | 0.5672 | 0.662 | 5 | 0.111 | 0.159 | 12 | 0.057 | 0.180 | 12 | 0.054 | 0.135 | 12 |
| | 0.013 | 0.051 | 11 | 0.448 | 0.347 | 1 | 0.3877 | 0.444 | 2 | 0.081 | 0.108 | 2 | 0.046 | 0.147 | 2 | 0.048 | 0.115 | 5 |
| Low | 0.013 | 0.051 | 2 | 0.393 | 0.250 | 6 | 0.3562 | 0.406 | 11 | 0.079 | 0.106 | 6 | 0.019 | 0.058 | 11 | 0.038 | 0.079 | 11 |
| | 0.005 | 0.019 | 9 | 0.368 | 0.208 | 7 | 0.2012 | 0.218 | 9 | 0.056 | 0.068 | 11 | 0.010 | 0.031 | 9 | 0.034 | 0.067 | 2 |
| | 0.003 | 0.011 | 6 | 0.287 | 0.068 | 2 | 0.0462 | 0.031 | 6 | 0.035 | 0.033 | 9 | 0.008 | 0.024 | 6 | 0.025 | 0.036 | 9 |
| Lowest | 0.000 | 0.000 | 1 | 0.248 | 0.000 | 5 | 0.0209 | 0.000 | 1 | 0.016 | 0.000 | 1 | 0.001 | 0.000 | 1 | 0.015 | 0.000 | 1 |

| Category | CLO Cent. Mea. | CLO Norm. value | CLO Sorted nodes | BET Cent. Mea. | BET Norm. value | BET Sorted nodes | EIV Cent. Mea. | EIV Norm. value | EIV Sorted nodes | HUB Cent. Mea. | HUB Norm. value | HUB Sorted nodes | AUT Cent. Mea. | AUT Norm. value | AUT Sorted nodes | PAR Cent. Mea. | PAR Norm. value | PAR Sorted nodes |
|---|---|---|---|---|---|---|---|---|---|---|---|---|---|---|---|---|---|---|
| Highest | 0.051 | 1.000 | 7 | 66.000 | 1.000 | 7 | 1.000 | 1.000 | 1 | 1.000 | 1.000 | 1 | 1.000 | 1.000 | 9 | 0.139 | 1.000 | 1 |
| | 0.049 | 0.955 | 5 | 56.000 | 0.848 | 5 | 0.737 | 0.682 | 9 | 0.093 | 0.091 | 11 | 0.756 | 0.737 | 8 | 0.130 | 0.905 | 9 |
| High | 0.041 | 0.800 | 8 | 48.000 | 0.727 | 4 | 0.407 | 0.282 | 8 | 0.068 | 0.065 | 9 | 0.518 | 0.482 | 12 | 0.109 | 0.707 | 2 |
| | 0.036 | 0.699 | 4 | 36.000 | 0.545 | 3 | 0.361 | 0.226 | 11 | 0.046 | 0.044 | 6 | 0.387 | 0.341 | 7 | 0.102 | 0.635 | 3 |
| | 0.034 | 0.650 | 3 | 27.000 | 0.409 | 10 | 0.318 | 0.174 | 12 | 0.035 | 0.032 | 4 | 0.382 | 0.335 | 10 | 0.092 | 0.545 | 8 |
| Medium | 0.030 | 0.579 | 2 | 20.000 | 0.303 | 2 | 0.239 | 0.078 | 10 | 0.025 | 0.023 | 2 | 0.331 | 0.282 | 11 | 0.085 | 0.479 | 12 |
| | 0.018 | 0.335 | 9 | 17.000 | 0.258 | 8 | 0.212 | 0.046 | 5 | 0.022 | 0.019 | 12 | 0.300 | 0.248 | 5 | 0.069 | 0.323 | 5 |
| | 0.017 | 0.315 | 10 | 0.000 | 0.000 | 9 | 0.211 | 0.045 | 3 | 0.021 | 0.018 | 3 | 0.207 | 0.148 | 6 | 0.064 | 0.272 | 11 |
| Low | 0.015 | 0.278 | 12 | 0.000 | 0.000 | 12 | 0.204 | 0.036 | 7 | 0.017 | 0.015 | 10 | 0.182 | 0.121 | 4 | 0.063 | 0.262 | 10 |
| | 0.014 | 0.245 | 6 | 0.000 | 0.000 | 6 | 0.184 | 0.011 | 2 | 0.008 | 0.006 | 8 | 0.155 | 0.092 | 3 | 0.062 | 0.256 | 4 |
| | 0.011 | 0.194 | 11 | 0.000 | 0.000 | 11 | 0.183 | 0.011 | 6 | 0.005 | 0.003 | 5 | 0.088 | 0.020 | 2 | 0.049 | 0.130 | 7 |
| Lowest | 0.002 | 0.000 | 1 | 0.000 | 0.000 | 1 | 0.174 | 0.000 | 4 | 0.003 | 0.000 | 7 | 0.069 | 0.000 | 1 | 0.036 | 0.000 | 6 |

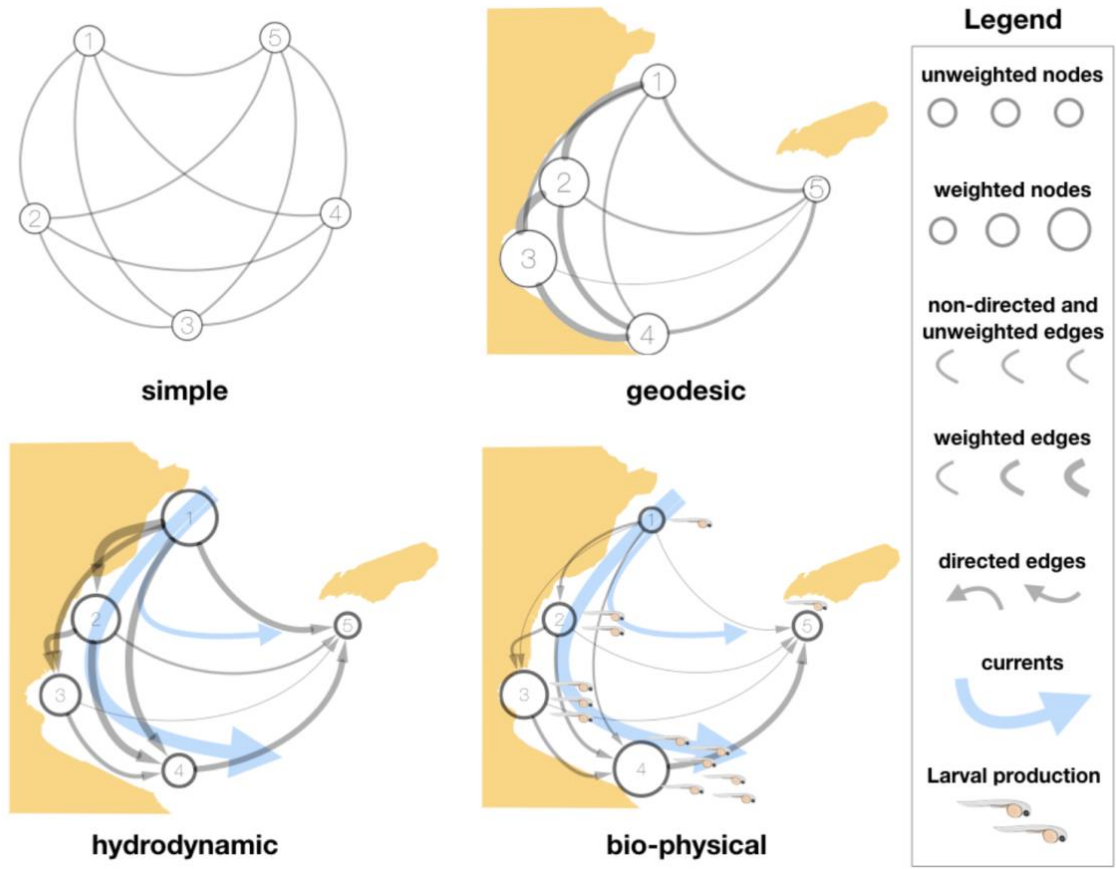

Figure S1: Four hypothetical graph networks of five nodes representing connected sites. Upper left panel shows a simple non-directed and unweighted graph network. Upper right panel shows a geodesic non-directed and weighted graph network, the number of nodes is the same, but now each node has assigned a geographical position. The importance of the nodes (weights) depends of the sum of all distances to other nodes. The strength of the linkages, showed as edges, depends on distance between node pairs. Lower left panel represents a hydrodynamic directed and weighted graph network, the geographical position of nodes is as in the geodesic network. The nodes and edges weights depend of the probability of connection determined by ocean currents. The directionality of ocean currents implies directionality in the linkages. This graph corresponds to a potential connectivity network. The lower right panel shows a bio-physical directed and weighted graph network. Here, the ocean currents exert influence on directionality and weighting of nodes and edges, but biological and ecological conditions determine the resulting importance of each node and linkage in the network. This graph corresponds to a realised larval connectivity network.

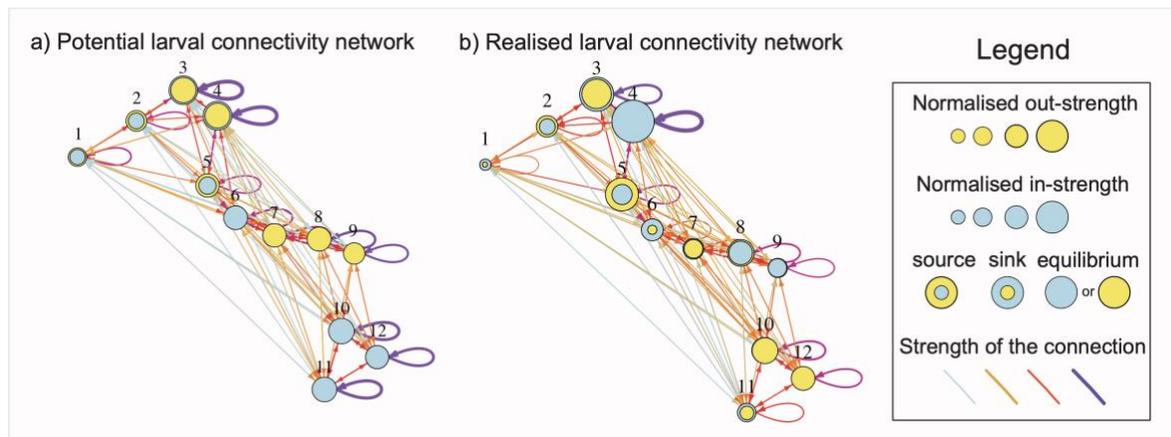

Figure S2: Graphs for the a) potential and b) realised larval connectivity networks. The nodes are symbolising Marine Protected Areas (MPAs) and Open Access Areas (OAAs) as in Fig. 1. The larval export and recruitment are showed as the normalised Out-Strength (yellow) and In-Strength (blue), respectively. When exportation is higher than recruitment the node represents a source site. When recruitment is higher than exportation the node represents a sink site. The strength of the connection is showed as coloured edges between node pairs.

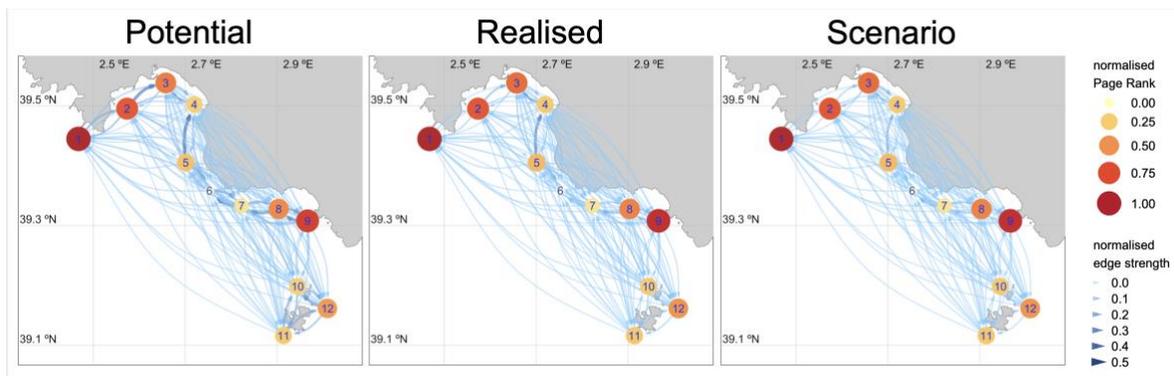

Figure S3: Page Rank for the potential, realised and scenario larval connectivity networks. Node size and colour symbolise Page Rank and edge width and colour symbolise normalised edge strength. Note that Page Rank for the nodes show no difference between the different networks, suggesting that this measure of centrality is only of limited use in the study of larval connectivity.

**Supplementary Methods: Individual Based Model for Total Egg production Method**

The hydrodynamic directed network corresponds to a graphical representation of a potential larval connectivity matrix (see main document). In contrast, a realized larval transport directed network corresponds to a graphical representation of a realized larval connectivity matrix. We calculated the realized larval connectivity matrix weighting the potential larval connectivity matrix (probability of connection due to hydrodynamics) with the spatial variability in the abundance of eggs produced and released, derived from available reproductive output information for the painted comber *Serranus scriba* (Alonso-Fernández et al., 2011; Alos et al., 2013; 2014). To estimate the total number of eggs released every week in each of the nodes we developed a daily egg production probabilistic Individual Based Model (egg-IBM). The egg-IBM was based on a simulation model of the total number of eggs released considering the empirical number of individuals (abundance) and the size structure of each node, and the reproductive parameters estimated in two Bayesian models using empirical data: a spawning probability model (model 1) and the number of eggs released by each spawning individual in function of the Julian day and the size of the fish (total length, mm) (model 2).

In the first Bayesian Model (model 1), we estimated the parameters of a logistic regression model of being mature or not against the Julian date and fish size using a data-set published in Alonso-Fernandez et al. (2011). Accordingly, 753 individuals of *S. scriba* were sampled in different locations of Mallorca island over the years 2006 and 2007, their gonads dissected and their maturity stage (spawning vs. non-spawning) histologically assessed (Alonso-Fernández et al., 2011). The parameters of the regression model were estimated using a Bayesian approach considering flat and uninformative priors. Interestingly, the single effect of size was significant (the Bayesian Credibility Interval-BCI didn't overlap zero) suggesting the larger the individual, the longer the spawning season. The posterior distributions of the model parameters were further used to predict the probability of a simulated individual being in spawning stage or not.

In the second Bayesian model (model 2), we estimated the parameters of an exponential model fitting the number of eggs released by an individual every day (*S. scriba* is a daily batch spawner, Alos et al., 2013) in function of their fish size and the spatial location of the node. Thus, the batch fecundity (number of hydrated eggs) of 129 individuals sampled across the 12 nodes of the network in May 2007 was estimated. Alos et al. (2014) have shown that *S. scriba* displays significant differences in reproductive investment of isolated sub-populations on small spatial scales in the study area. Therefore, in order to achieve a more realistic realized larval connectivity matrix for a MPA-OAA mixed network, we have estimated the spatially structured egg production of *S. scriba* for all nodes considered in our matrix: nodes for the National Park of Cabrera (n=19), Palma Bay (n=27) and the rest (n=83) were aggregated. The parameters of the three spatial regression models were estimated using a Bayesian approach considering the same flat and uninformative prior

structure. The posterior distributions of the model parameters were further used to predict the number of eggs released by an individual with a given fish size.

The estimation of daily egg production (DEP) for each node or sub-population was based on the modification of classical equations to estimate indices of reproductive potential (Morgan, 2008) with the following formula:

$$DEP = \sum_{i=1}^{n} N_i \times BF_i \times SF_i d$$

where $N_i$ is the number of individuals at length $i$, $BF_i$ is the individual batch fecundity at length $i$ and $SF_i d$ is the fraction of individuals in actively spawning phase at length $i$ in day $d$ along the year.

The egg-IBM was then run considering the parameters estimated in Bayesian models 1 and 2 to estimate the total number of eggs released in each node for a given year (here, 2009). Accordingly, we first estimated the total number of individuals in each of the nodes (abundance of fish in each node). Briefly, we used an experimental trawl (Alos et al., 2014) to obtain a measure of density (number of individuals) per unit of area of suited habitat (*Posidonia oceanica* seagrass) in 2006. In the National Park of Cabrera, it was not allowed to sample using this gear, and we used a relationship between hook-and-line catch rates and experimental trawl relationship extracted from the other nodes to predict the number of individuals in the National Park. Then, we used a combination of the data from the experimental trawl and hook-an-line data to produce an estimate of the size structure of the fish in each node. Despite both density and size structure were estimates, this approach produced a realistic demographic and high representative measures to be incorporated in our egg-IBM.

The egg-IBM simulated the daily number of eggs produced by the whole set of individuals from each node (with their size accuring to the size structure of the node) for the whole year 2009 according to the probability of spawning (model 1), and if their spawn, how many eggs released every day according to the parameters estimated in model 2. As the connectivity matrices mentioned above were at a weekly scale, the total number produced by each individual was transformed to weekly-eggs produced by each node by aggregating daily-eggs produced every week in each node. While our model was fundamentally deterministic, we accounted for uncertainty in the two models' parameters by running our egg-IBM until 1,000 times, each time re-sampling from the posterior distribution of the model parameters (Bayesian Credibility Intervals) and the empirical distribution of fish density and size structure (mean and variance) to obtain 1,000 values of the total eggs produced by each node every week. The final number of eggs generated every week where multiplied by the connectivity matrix described before to determine the fate of the total number of eggs released in each node and compute the realized larval connectivity matrix weighed by the potential larval connectivity matrix. The realised larval connectivity network presented in the manuscript is the average of 240 matrices, each corresponding to one week from March 21 to August 29 (30 weeks per year) for 8 years from 2000 to 2009. This graph network is a snapshot of the current state of *S. scriba* population living within the mixed MPAs-OAAs network and subjected to fishing pressure. In consequence, the egg

abundance is a function of the level of protection, the adult reproductive population characteristics and the available sea grass / rocky bottom habitat.



\end{document}